\documentclass[a4paper,fleqn,usenatbib]{mnras}

\usepackage{newtxtext,newtxmath}

\usepackage[T1]{fontenc}
\usepackage{ae,aecompl}

\usepackage{graphicx}	
\usepackage{amsmath}	
\usepackage{amssymb}	

\newcommand{\atlas}{\textsc{ATLAS$^\mathrm{3D}$}}
\newcommand{\gadget}{\textsc{Gadget-3}}
\newcommand{\mappings}{\textsc{MAPPINGS III}}
\newcommand{\skirt}{\textsc{SKIRT}}
\newcommand{\galfit}{\textsc{Galfit}}
\newcommand{\kinemetry}{\textsc{KINEMETRY}}

\title[The fate of the Antennae galaxies]{The fate of the Antennae galaxies}

\author[N. Lah\'{e}n et al.]{Natalia Lah\'{e}n$^{1}$\thanks{E-mail: natalia.lahen@helsinki.fi}, 
Peter H. Johansson$^{1}$,
Antti Rantala$^{1}$,
Thorsten Naab$^{2}$, 
\newauthor{Matteo Frigo$^{2}$
}
\\
$^{1}$Department of Physics, University of Helsinki, Gustaf 
H$\ddot{a}$llstr$\ddot{o}$min katu 2a, FI-00014 Helsinki, Finland, \\
$^{2}$Max Planck Institute for Astrophysics, Karl-Schwarzschild-Str. 1, D-85748, Garching, Germany \\
}

\date{Accepted XXX. Received YYY; in original form ZZZ}

\pubyear{2017}

\begin{document}
\label{firstpage}
\pagerange{\pageref{firstpage}--\pageref{lastpage}}
\maketitle

\begin{abstract}

We present a high-resolution smoothed particle hydrodynamics simulation of the Antennae galaxies (NGC 4038/4039) and follow the evolution
$3$ Gyrs beyond the final coalescence. The simulation includes metallicity dependent cooling, star formation, and both stellar feedback and chemical
enrichment. The simulated best-match Antennae reproduces well both the observed morphology and
the off-nuclear starburst. We also produce for the first time a simulated two-dimensional metallicity 
map of the Antennae and find good agreement with the observed metallicity of off-nuclear stellar clusters, 
however the nuclear metallicities are overproduced by $\sim 0.5$ dex. 
Using the radiative transfer code \skirt\ we produce multi-wavelength observations of both the Antennae and the 
merger remnant. The $1$ Gyr old remnant is well fitted with a S\'{e}rsic profile of $n=7.07$, and with an $r$-band effective radius of
$r_{\rm e}= 1.6$ kpc and velocity dispersion of $\sigma_{\rm e}=180$ km$/$s the remnant is located on the fundamental plane of 
early-type galaxies (ETGs). The initially blue Antennae 
remnant evolves onto the red sequence after $\sim 2.5$ Gyr of secular evolution. The remnant would be classified as a fast rotator, 
as the specific angular momentum evolves from $\lambda_{\rm{Re}}\approx0.11$ to $\lambda_{\rm{Re}}\approx0.14$ during its evolution.
The remnant shows ordered rotation and a double peaked maximum in the mean 2D line-of-sight velocity. 
These kinematical features are relatively common among local ETGs and we specifically identify  three local 
ETGs (NGC 3226, NGC 3379 and NGC 4494) in the \atlas\ sample,  whose photometric and kinematic properties most resemble the Antennae remnant. 

\end{abstract}




\section{Introduction}

In the early days of galaxy studies early-type galaxies (ETGs) were considered to be a homogeneous class of galaxies characterised by their nearly 
spherical shapes, old stellar populations and the lack of cold gas and on-going star formation 
\citep{1976ApJ...204..668F,1977ApJ...218..333K,1987ApJ...313...59D}. However, it is now well
established that the fine structural and kinematic properties of elliptical galaxies correlate with their absolute magnitudes 
\citep{1983ApJ...266...41D,1994MNRAS.269..785B,1996ApJ...464L.119K,2005ApJ...621..673T}.  
In general, brighter ellipticals $(M_{\rm B} \lesssim -21.5)$ typically have boxy isophotes, cored central surface brightness profiles and
often show radio and X-ray emission (e.g. \citealt{1997AJ....114.1771F,2009ApJS..182..216K,2016ARA&A..54..597C}). Bright massive ellipticals are also typically 
slowly rotating $(v/\sigma<0.1)$ moderately triaxial systems, with often peculiar properties such as kinematic twists and kinematically 
decoupled components (e.g. \citealt{2007MNRAS.379..418C,2011MNRAS.414..888E,2011MNRAS.414.2923K}). Fainter 
early-type galaxies $(-21.5 \lesssim M_{\rm B} \lesssim -18)$, on the other hand, are in general oblate and 
have typically more discy isophotes, cuspy central surface brightness profiles and typically show no significant X-ray emission (e.g. 
\citealt{2009ApJS..182..216K,2012ApJS..198....2K}).
These galaxies are also rotating faster $(v/\sigma\sim 1)$ and are intrinsically more anisotropic systems than slowly-rotating 
ellipticals (e.g. \citealt{2007MNRAS.379..418C,2011MNRAS.414..888E,2016ARA&A..54..597C}).

The first models attempting to describe the formation of early-type galaxies invoked a monolithic collapse \citep{1962ApJ...136..748E}, whereas
more modern galaxy formation models are steeped in the hierarchical picture of structure formation, in which galaxies grow bottom-up 
through mergers and gas accretion \citep{1978MNRAS.183..341W}. In recent years a consensus has tentatively emerged in which massive, slowly-rotating 
early-type galaxies are believed to have assembled through a two-stage process. In this picture the early assembly is dominated by rapid in situ star 
formation fuelled by cold gas flows and hierarchical merging of multiple star-bursting progenitors, whereas the later growth at lower redshifts, below
$z\lesssim 2-3$ is dominated by a more quiescent phase of accretion of stars formed mainly in progenitors outside the main galaxy (e.g. 
\citealt{2009ApJ...699L.178N,2010ApJ...725.2312O,2011ApJ...736...88F,2012ApJ...754..115J,2015MNRAS.449..361W,2017ARA&A..55...59N}).

On the other hand, some of the lower mass ETGs might have formed instead in the merger scenario, in which elliptical galaxies form as result of the merger 
of two disc galaxies \citep{1977egsp.conf..401T,1983MNRAS.205.1009N,1992ARA&A..30..705B,2017ARA&A..55...59N}. Much of the 
theoretical understanding of the merger scenario has been gained using idealised isolated merger 
simulations, with the first studies employing purely collisionless models (e.g. \citealt{1978MNRAS.184..185W,1980ApJ...236...43A}); 
later studies included also the gaseous component (e.g. \citealt{1989ApJS...70..419H,1996ApJ...471..115B,2006MNRAS.372..839N}) 
and increasingly sophisticated models for gas cooling, star formation and the feedback from both supernovae and supermassive black holes 
 (e.g. \citealt{2005MNRAS.361..776S,2006MNRAS.373.1013C,2009ApJ...690..802J,2009ApJ...707L.184J,2010ApJ...720L.149T,2013MNRAS.430.1901H,2014MNRAS.442.1992H}). 
These idealised merger simulations demonstrated that kinematically cold discs can be transformed into hot spheroidal systems with elliptical-like
properties (e.g. \citealt{1983MNRAS.205.1009N,1992ARA&A..30..705B}). The merger mass-ratio was 
found to have a significant impact on the merger remnants, with more unequal-mass mergers
resulting in remnants with faster rotation and more discy isophotes, whereas more equal-mass mergers would result in slower rotation and somewhat more boxy
isophotes (e.g. \citealt{2003ApJ...597..893N,2005A&A...437...69B,2006ApJ...650..791C,2009MNRAS.397.1202J,2010ApJ...723..818H,2011MNRAS.416.1654B,2017ARA&A..55...59N}).

Although the merging of disc galaxies might not be the dominant formation mechanism for the majority 
of ETGs (e.g. \citealt{2009ApJ...690.1452N}), they are nevertheless observed in the nearby Universe. 
The Antennae galaxies (NGC 4038/4039) are the nearest and arguably best studied example of an ongoing 
major merger of two gas-rich spiral galaxies. This system contains two clearly visible, still distinct 
discs, together with the signature feature of a beautiful pair of narrow elongated tidal tails extending to a projected 
size of $~20'$, which corresponds to $\sim 100$ kpc at an estimated distance of $20$ Mpc. The exact distance to 
the Antennae system is still surprisingly uncertain. The often quoted value of $\sim 20$ Mpc has been derived from the systemic recession 
velocity and an adopted Hubble constant of $H_{0}=72$ km s$^{-1}$ Mpc$^{-1}$ with or without corrections
for deviations of the Hubble flow from linearity due to various attractors \citep{1999AJ....118.1551W,2008AJ....136.1482S}. Significantly 
smaller distances of $13.3$--$13.8$ Mpc have also been determined based on the photometry of the tip of the red giant branch 
\citep{2004AJ....127..660S, 2008ApJ...678..179S}. However, recent observations of the type Ia supernova 2007sr and Cepheid variables in the Antennae seem to agree
on a distance of $\sim 22$ Mpc in relatively good agreement with the traditional distance 
\citep{2008ApJ...678..179S,2011ApJ...730..119R, 2013AJ....146...86T, 2013MNRAS.434.2866F}, although distances in excess of $\sim 25$ Mpc have also been advocated
based on the brightness and decay time of the light curve of the supernova 2007sr \citep{2008MNRAS.388..487N}.

The proximity of the Antennae galaxies and the availability of detailed high-quality observations covering the entire
spectrum from radio to X-ray (e.g. \citealt{1999AJ....118.1551W,2001ApJ...554.1035F,2004ApJS..154..193W,2005ApJ...619L..87H,2006ApJS..166..211Z,2015ApJ...815..103B}), 
including recent ALMA observations \citep{2012A&A...538L...9H,2014ApJ...795..156W} makes this system an ideal laboratory for 
understanding the physics of merger-induced starbursts through comparison with high-resolution simulations. In galaxy mergers the interaction
typically induces strong spiral arms and extended tidal tails in the merging galaxies. The interaction also gives rise to central tidal torques that 
efficiently transport gas to the centres of the galaxies (\citealt{1991ApJ...370L..65B}), where the rapid build-up of dense gas 
typically results in 
central starbursts (\citealt{1996ApJ...464..641M}). However, the Antennae system is particularly interesting, as currently the majority of the star formation 
activity is not observed in the nuclei, but rather in a dusty overlap region between the merging galactic 
discs \citep{1998A&A...333L...1M,2004ApJS..154..193W,2012ApJ...745...65U}, making this system a rare extranuclear starburst. 

The Antennae galaxies have been a favourite and natural target for numerical simulations, with the very first simulation by \citet{1972ApJ...178..623T} already 
reproducing the correct trends in the morphology of the tidal arms. Refinements that included self-consistent multi-component galaxy models consisting 
of a bulge, disc and dark matter halo \citep{1988ApJ...331..699B} and finally models that also included gas and star formation 
\citep{1993ApJ...418...82M} also reproduced the general morphology 
of the Antennae well. The introduction of adaptive-mesh refinement techniques together 
with stellar feedback models have enabled a more detailed modelling of the
star formation histories of the Antennae galaxies \citep{2010ApJ...720L.149T,2015MNRAS.446.2038R}.    
However, the star formation in all of these simulations was mostly centrally concentrated, thus the observed 
elevated star formation rate in the overlap region was not reproduced. On the contrary, more recent simulations in which the time of 
best match is placed after the second encounter, shortly before the 
final coalescence, are able to produce a gas-rich starforming overlap region in the Antennae system
\citep{2008AN....329.1042K,2010ApJ...715L..88K,2011ApJ...734...11K,2013MNRAS.434..696K,2010ApJ...716.1438K}. This is also the orbital configuration adopted 
in this paper. 

In this paper we study the fate of the Antennae galaxies. The main motivation of this paper is two-fold. Firstly, we want to produce an accurate simulation
of the Antennae galaxies, a bona-fide representation of a specific real galaxy merger. This includes a detailed dynamical model, 
elaborate star formation and feedback models including metal enrichment, and the production of mock-images for a direct comparison with the 
observations. Secondly, we evolve the simulation past the best present-day match and into the future in order to study whether the resulting 
Antennae merger remnant will evolve into an low-mass discy early-type galaxy, as predicted by the merger scenario. Thus the aim is to test whether the 
early-type galaxy formation through galaxy mergers is a viable process in the local Universe and whether this process can explain the formation of some of the
observed local low-mass early-type galaxies. The Antennae galaxies provides us also with a prime example of a present-day major merger of
two spiral galaxies with significant star formation and their evolutionary state can be linked to the more luminous LIRGs and 
ULIRGs (Luminous/Ultra-luminous infrared galaxies) that are thought to be very gas-rich disc major mergers undergoing extreme star formation and morphological
transition to elliptical galaxies (e.g. \citealt{1998ApJ...498..579G,2006ApJ...638..745D,2006ApJ...651..835D,2017MNRAS.471.2059V}). 

We produce mock-images of the simulated Antennae merger and the dynamically relaxed merger remnant by combining stellar population synthesis 
and radiative transfer modelling \citep{2003MNRAS.344.1000B,2008ApJS..176..438G,2015A&C.....9...20C}.
The resulting photometric images of the Antennae and the merger remnant are compared to the present 
day Antennae and a sample of local early-type galaxies, respectively.
Our multi-wavelength radiative transfer analysis goes significantly beyond the radiative transfer modelling study presented in \citet{2013MNRAS.434..696K}, which found
a good match between their simulated Antennae and the far-infrared observations performed by the Herschel-PACS instrument \citep{2010A&A...518L..44K}.
In addition we also construct two-dimensional kinematic maps of the merger
remnant using \kinemetry\ methods \citep{2006MNRAS.366..787K,2014MNRAS.444.3357N}. Recently several observational surveys, 
such as \atlas\ (e.g. \citealt{2011MNRAS.413..813C,2011MNRAS.416.1654B}), CALIFA (e.g. \citealt{2012A&A...538A...8S,2014A&A...567A.132W}) and 
SAMI (e.g. \citealt{2015MNRAS.447.2857B}), that employ integral field units have provided detailed kinematic maps of local early-type galaxies. Here 
we study the kinematics of the future
Antennae remnant in order to determine whether a local look-alike exists. 

In \S 2 we discuss the details of our simulations, including the setting up of the initial conditions. The necessary post-processing tools that are required for the
analysis are described in \S 3. In \S 4 we show the simulated Antennae galaxies at the time of the best match and compare the derived morphology, star formation
rate and spectral energy distribution to observations. We discuss the intrinsic properties, such as the shape, metallicity and velocity structure, 
of the simulated merger remnant in \S 5. The photometric light profile and colour of the future Antennae remnant is described in \S 6. In \S 7 we construct 
kinematic maps of the Antennae remnant and derive its rotational properties. In \S 8 we search for a best match to our simulated Antennae remnant 
in the observational \atlas\ catalogue 
and demonstrate that the future Antennae remnant will have properties reminiscent of local early-type galaxies. Finally in \S 9 we summarise our 
results and present our conclusions. 

\section{Simulations}

\subsection{Simulation code}

Our simulations are run using the N-body smoothed particle hydrodynamics (SPH) code 
\gadget\ \citep{2005MNRAS.364.1105S} using specifically the SPHGal version of the code 
\citep{2014MNRAS.443.1173H,2017MNRAS.468..751E}. SPHGal includes several improvements of the hydrodynamics 
implementation, which we briefly review here. The spline kernel of standard \gadget\
has been replaced by a Wendland $C^4$ kernel \citep{2012MNRAS.425.1068D} and the number of 
neighbours in the SPH smoothing kernel has been increased to 100. The code uses the pressure-entropy formulation 
of SPH \citep{2013ApJ...768...44S,2013MNRAS.428.2840H}, a higher order estimate of velocity gradients \citep{2010MNRAS.408..669C}, 
a modified artificial viscosity switch with a modified strong limiter \citep{1997JCoPh.136...41M,2010MNRAS.408..669C}, and artificial 
conduction of thermal energy \citep{2008JCoPh.22710040P}. Finally, SPHGal employs a time step limiter which activates particles 
receiving feedback energy and keeps the time steps of neighbouring particles within a factor of four of each other. 
Together these changes reduce the numerical artefacts present in the fluid mixing that afflicted the  
original \gadget\ version, and also considerably improve the convergence rate of the SPH calculation. 

\subsection{Subgrid models}

Astrophysical processes such as metallicity-dependent cooling, star formation, 
stellar feedback, as well as metal production and metal diffusion are followed using 
subresolution models originally developed by  \citet{2005MNRAS.364..552S, 2006MNRAS.371.1125S}
and later extended by \citet{2013MNRAS.434.3142A}.
The gas in the simulation cools with a rate dependent on the current gas temperature, density and 
metal abundances. The cooling rates are adopted from \citet{2009MNRAS.393...99W} \citep[see also][]{2013MNRAS.434.3142A} for an 
optically thin gas in ionisation equilibrium and the cooling rates are calculated on a element-by-element basis taking
into account the effects of uniform redshift-dependent ionising UV/X-ray background \citet{2001cghr.confE..64H}, with 
the background calculated at $z=0$ for our case. The cooling rates are tabulated for temperatures between $10^{2}$ K $\le T \le 10^{9}$ K 
but below $T=10^4$ K the cooling becomes inefficient due to the lack of molecular cooling in the current model. 

Gas particles with densities $\rho_\mathrm{g}$ above the threshold density of 
$\rho_\mathrm{crit}=1.6\times10^{-23}$ $\mathrm{g/cm^{3}}$ corresponding to
$n_{\rm H}=10$ cm$^{-3}$, and with temperatures below $T< 12000$ K, are eligible for star formation. The probability 
for a gas particle to turn into a stellar particle is given by $1-e^{-p}$, where the coefficient $p$ 
is defined using the local dynamical time $t_\mathrm{dyn}$ as 
\begin{equation}
 p=\epsilon_\mathrm{SFR}\frac{\Delta t}{t_\mathrm{dyn}}
=\epsilon_\mathrm{SFR}\Delta t \sqrt{4\pi G\rho_\mathrm{g}},
\end{equation}
where $\Delta t$ is the length of the current simulation time step and the star formation 
efficiency is set to $\epsilon_\mathrm{SFR}=0.02$. 

Stellar particles couple to the surrounding gaseous particles through
heating and mass transfer with kinetic
feedback and chemical enrichment from
supernovae type Ia (SNIa), type II (SNII) and asymptotic
giant branch stars (AGB). The stellar and gaseous 
abundances of 11 individual elements (H, He, C, Mg, O, Fe, Si, 
N, Ne, S, Ca) evolve based on models of chemical release rates from \citet{1999ApJS..125..439I} for SNIa, 
\citet{1995ApJS..101..181W} for SNII and \citet{2010MNRAS.403.1413K} for AGB stars, respectively. 
Variations in the metallicity between neighbouring gaseous particles are smoothed using
the diffusion implementation of \citet{2013MNRAS.434.3142A} which employs the definition for the diffusion
coefficient from \citet{2010MNRAS.407.1581S}, with the diffusion coefficient pre-factor set to $0.05$. Small
changes in the diffusion coefficient are not expected to qualitatively change the derived metallicity profiles.

Stellar feedback begins with SNII explosions $3$ Myr after the formation of a stellar particle,
followed by SNIa and AGB feedback from $50$ Myr onwards. 
SNII feedback is released only once, whereas the SNIa and AGB feedback are 
released continuously from each stellar particle individually in $50$ Myr intervals for a total time period of $10$ Gyr.
The particle-by-particle SNIa/AGB ejecta release rates decay proportionally to 
$t^{-1}$ \citep{2011MNRAS.412.1508M}, with a total release 
of 2 SNIa per formed 1000 $M_{\sun}$. Each stellar particle flagged to give SN feedback releases an amount of energy equalling
\begin{equation}
 E_\mathrm{SN}=\frac{1}{2}m_\mathrm{eject} v_\mathrm{SN}^2
\end{equation}
into the interstellar matter, where $m_\mathrm{eject}$ is the mass of the SN ejecta and 
$v_\mathrm{SN}=4000$ km$/$s is the SN ejecta velocity. 
The metallicity dependent ejecta mass is distributed amongst the $10$
nearest gas particles weighted by the corresponding SPH smoothing kernel
of the stellar particle. 

The thermal and kinetic feedback are distributed to the gas particles within the smoothing kernel
in three phases, depending on the distance of the receiving particle 
from the SN (see \citealt{1988ApJ...334..252C} for a review of the evolutionary 
phases of SN remnants). 
The transition radii between the different phases depend on the ejecta mass and 
velocity, as well as the local gas density, see \citet{2017ApJ...836..204N} for a detailed 
description of the transition radii and the distribution of the SN energy.
The three phases are the free expansion (FE) phase with momentum conservation, the adiabatic Sedov-Taylor (ST) phase with 
heating \citep{1950RSPSA.201..159T, 1959sdmm.book.....S}, 
and the snowplow (SP) phase with efficient radiative cooling 
\citep{1977ApJ...218..148M, 1998ApJ...500..342B}. Gas 
particles in the FE region, i.e. closest to the SN, receive only kinetic 
energy. The transition between the FE and ST phases is set at the radius  
within which the shocked ISM mass exceeds the SN ejecta mass.
Gas particles in the ST region receive $70\%$ of the injected energy as 
thermal and $30\%$ as kinetic energy, as the shock ejecta are slowed down by the 
local ISM. In the outermost region the velocity of the SN ejecta decreases 
further, finally dispersing the shock. In the SP region gas is allowed to cool, 
and the energy is distributed similarly to the ST phase, except the amount of 
injected energy is reduced as a function of distance from the SN, leading to 
radiative loss of energy. Energy losses in the SP phase are steeper for thermal
energy than the kinetic energy, thus at the outermost radii the SN feedback 
is predominantly dominated by the kinetic component (see e.g. \citealt{2017ApJ...836..204N}).

The AGB energy and enrichment are distributed in the same fashion 
as the SN feedback but only in the FE phase, with a much lower wind velocity 
of $25$ km$/$s and lower yields adopted from \citet{2010MNRAS.403.1413K}. The masses 
of the putative supermassive black holes in the Antennae galaxies are currently unknown. Thus, the
simulation presented in this paper does not include supermassive black holes.

\subsection{Initial conditions}

In order to study the evolution of the Antennae merger in a controlled environment,
we run a high resolution major merger of two idealised disc galaxies. The 
disc galaxies represent the progenitor galaxies NGC 4038 and NGC 4039 in the ongoing merger.
We set up the initial conditions following the parameters of \citet{2010ApJ...715L..88K} (see Table \ref{tab:ic_parameters})
with additional refinements concerning the age and metallicity distribution of the stellar and gaseous particles at the start
of the simulation (see Sec. \ref{section:age_description} and \ref{section:metallicity_description}).

Both of the progenitor galaxies have virial masses of 
$M_\mathrm{vir}=5.52\times 10^{11} \ M_{\sun}$. Each progenitor consists of a 
stellar bulge, a stellar disc and a gaseous disc embedded in a dark matter halo.  Both dark matter haloes have Hernquist 
density profiles \citep{1990ApJ...356..359H} with NFW density profile equivalent 
concentration parameters of $c=15$ (for a connection between the Hernquist profile and the NFW
profile see \citealt{2005MNRAS.361..776S}). The
dimensionless spin-parameters \citep[see e.g.][]{2001ApJ...555..240B} are set to $\lambda_{4038}=0.1$ and 
$\lambda_{4039}=0.07$. The 
disc component of each galaxy, with a total mass of $M_\mathrm{disc}=4.14\times 
10^{10}$ M$_{\sun}$, receives a fraction of the angular momentum of the halo 
given by the disc mass fraction $m_\mathrm{d}=0.075$. Each disc has a gas 
fraction of $f_\mathrm{g}=0.2$, with the remainder of the disc consisting of stars. The 
rotationally supported stellar discs have scale lengths of
$r_\mathrm{d,4038}=6.28$ kpc and $r_\mathrm{d,4039}=4.12$ kpc and vertical scale heights 
of $z_\mathrm{d,4038}=1.26$ kpc and $z_\mathrm{d,4039}=0.82$ kpc, set as 
$20\%$ of the respective scale lengths. The gaseous discs have the same scale lengths 
as the stellar discs, and scale heights iterated through the requirement for 
hydrostatic equilibrium \citep{2005MNRAS.361..776S}. The stellar bulges with bulge-to-disc 
mass ratios of $1/3$ are set up also using the Hernquist density profile with scale 
lengths of $r_\mathrm{bulge}=0.2r_\mathrm{disc}$, resulting in a total baryon 
fraction of $f_\mathrm{b}=0.1$ for the progenitor galaxies. 

Each galaxy consists of $8.28\times10^{6}$ particles in total, with 
$2.76\times10^{6}$ dark matter particles, $3.31\times10^{6}$ stellar disc 
particles, $1.38\times10^{6}$ bulge particles and $8.3\times10^{5}$ gaseous 
particles. Baryonic matter is thus represented by macroparticles with an initial 
mass resolution of $m_\mathrm{b}= 1.0 \times 10^4 \ M_{\sun}$, whereas the dark matter 
has a mass resolution of $m_\mathrm{DM}=1.8\times10^5 \ M_{\sun}$. 
The mass resolution is therefore increased by a factor of $\sim 7$ compared to the one originally
adopted in \citet{2010ApJ...715L..88K}. We checked for convergence by running the simulation at 
both the original \citet{2010ApJ...715L..88K} resolution and at an intermediate resolution, and found 
similar results by adjusting only the time of best match,
i.e. which particular snapshot represents best the observations.
The gravitational softening lengths are set at $\epsilon_\mathrm{b}=13$ pc for the baryonic components
and $\epsilon_\mathrm{DM}=55$ pc for the dark matter, respectively. 
The simulation was run for a total of
$4.3$ Gyr, which corresponds to an evolution of $3$ Gyr past the final coalescence of the galaxies, which takes place 
at approximately $t=1.3$ Gyr.

\begin{table}
 \caption{Parameters for the progenitor disc galaxies.}
 \label{tab:ic_parameters}
 \begin{minipage}{6cm}
 \begin{tabular}{lcc}
  \hline
  Property & NGC 4038 & NGC 4039\\
  \hline
  $M_\mathrm{vir}$\footnote{Masses in $10^{10}$ M$_{\sun}$}& $55.2$ & 
$55.2$  \\[2pt]
  $M_\mathrm{disc, stellar}$ & $3.3$ & $3.3$ \\[2pt]
  $M_\mathrm{disc, gaseous}$ & $0.8$ & $0.8$ \\[2pt]
  $M_\mathrm{bulge}$ & $1.4$ & $1.4$ \\[2pt]
  \hline
  $r_\mathrm{disc}$\footnote{Scale radii and heights in kpc} & $6.28$ & $4.12$ 
\\[2pt]
  $z_\mathrm{disc}$ & $1.26$ & $0.82$ \\[2pt]
  $r_\mathrm{bulge}$ & $1.26$ & $0.82$ \\[2pt]
  $c$ 	& $15$ & $15$ \\[2pt]
  $\lambda$ & $0.10$ & $0.07$ \\[2pt]
  \hline
  $i$ \footnote{Angles in degrees} & $60$ & $60$ \\[2pt]
  $\omega$ & $30$ & $60$ \\[2pt]
  \hline
 \end{tabular}
 \end{minipage}
\end{table}

The galaxies are set on nearly parabolic orbits with an initial separation of one 
virial radius, $r_\mathrm{sep}=r_\mathrm{vir}=168$ kpc and the pericentric 
separation set to the sum of their disc scale lengths, 
$r_\mathrm{p}=r_\mathrm{d,4038}+r_\mathrm{d,4038}=10.4$ kpc. The galaxies are 
rotated with disc inclinations of $i_{4038}=60^{\circ}$ and $i_{4039}=60^{\circ}$ 
and arguments of pericentre of $\omega_{4038}=30^{\circ}$ and 
$\omega_{4039}=60^{\circ}$. This orbital configuration has been shown to reproduce a good match to 
the spatial distribution and line-of-sight kinematics of the baryonic matter in 
the observed Antennae counterparts by \citet{2010ApJ...715L..88K}.

\subsection{Initial distribution of stellar ages}\label{section:age_description}

The simulation is started $1.2$ Gyr in the past corresponding to a redshift of $z\sim0.1$. 
This results in the best match between the observed and 
simulated Antennae at $z=0$ corresponding to a $13.7$ Gyr old universe.
In order to capture the feedback from stars at 
the start of the simulation we assign cosmic formation times and 
metallicities to the stellar particles present at the beginning of the simulation.
The selected ages and metallicities are derived from observationally supported 
star formation histories and metallicity distributions separately for the bulge, and the 
stellar and gaseous discs. 

The stellar mass in the bulges of disc galaxies in the local universe consists predominantly 
of old stars.
To obtain the distribution of cosmic formation times of the 
stellar particles in the bulge we apply an exponentially decaying star 
formation rate (SFR)
\begin{equation}
 \mathrm{SFR}_\mathrm{b}(t) = C e^{-(t-t_{\rm{0}})/\tau},
\end{equation}
where $\tau=1$ Gyr is the star formation timescale, $t$ is the cosmic time, $t_{\rm{0}}$ is the starting time of the simulation
and $C$ is a normalisation factor. We set the cosmic time at which SF begins to $500$ Myr and also assume 
that the contribution from SF in the bulge is negligible at the start of the simulation ($z\sim0.1$), reminiscent of the bulges in local disc galaxies 
(e.g. \citealt{2017A&A...597A..97M}).

The formation times of the stellar particles in the disc are acquired by 
assuming a linearly decaying star formation 
history with an initial SFR at $z\sim 0.1$ similar to a local star-forming stellar disc. The initial
SFR values are obtained through an iterative process where we first make an 
initial estimate, set up the galaxies and run them in isolation and adjust this estimate 
until we have a continuous SFR transitioning from the pre-set values 
to the actual SFR at the start of the simulation. However, the SFR at the beginning of the simulation
is mainly governed by the gas content of the galaxy, and with the gas fraction of $f_{\rm g}=0.2$ 
we are able to produce stable star-forming discs with initial SFR values of
$0.5$ M$_{\sun}/$yr for the extended NGC
4038 and $1.5$ M$_{\sun}/$yr for the more compact NGC 4039.

The ages of the individual stellar particles are obtained by integrating the time-dependent SFR
and requiring that all the stellar mass in both the bulge and the disc have 
formed before the start of the simulation. The ages of the 
stellar particles are then randomly sampled from the cumulative distribution functions
for both the bulge and the disc components, with an assigned age scatter of $\Delta t=100$ Myr. 

\subsection{Initial distribution of metals}\label{section:metallicity_description}

The initial element abundances of the stellar and gaseous particles are set 
by assigning uniform, radially decaying metallicity profiles.
The bulge, the stellar disc and the gaseous disc are set 
up with equal radial gradients. 
We use the Milky Way as a reference, as our progenitor 
galaxies have stellar masses quite similar to the Milky Way, for which stellar mass estimates range between 
$\sim 4$--$6.4\times 10^{10}$ M$_{\sun}$ (e.g. \citealt{2002ApJ...573..597K, 2011MNRAS.414.2446M}).
Moreover, our progenitor galaxies lie in the intermediate to massive end of the
stellar mass --  gas metallicity relation \citep{2004ApJ...613..898T}, where slight deviations
in stellar mass only result in small changes in metallicity.

The radial abundance profile of each metal $X$ is linear in dex-units, and is set as
\begin{equation}
 12+\log\left(\frac{X}{\mathrm{H}}\right) \equiv 
\left[\frac{X}{\mathrm{H}}\right]=k(r_s-r)+\left[\frac{X}{\mathrm{H}}\right]_{
r_s},
\end{equation}
where $k$ is the metallicity gradient, $r_s$ is the disc scale radius, 
$r$ is the distance from the galactic centre and 
$[X/\mathrm{H}]_{r_s}$ is the normalisation abundance at 
$r_s$. The observed oxygen gradient of $0.0585$ dex$/$kpc for the Milky Way from 
\citet{1994ApJ...420...87Z} is used as a reference, and the gradients in the two 
galaxies are scaled by keeping the integrated metal abundance equal to the Milky Way 
abundance. For our progenitor galaxies with slightly smaller stellar masses than the Milky Way,
this results in mildly steeper gradients of $k_\mathrm{4038}=0.0612$ dex$/$kpc and 
$k_\mathrm{4039}=0.0594$ dex$/$kpc for NGC 4038 and NGC 4039, respectively. Abundances of 
the nine metals at the respective scale radii are obtained from observations of 
the Milky Way at a radius of $3$ kpc according to \citet{1994A&A...291..757K} and 
\citet{1993PASP..105..327A} and are given in Table \ref{tab:ic_metals}.

\begin{table}
 \caption{Initial abundances of the nine individual metals at the disc scale 
radii (see Table \ref{tab:ic_parameters}), 
and the mean metal mass fractions in the stellar and gaseous components in
both of the progenitor galaxies.}
 \label{tab:ic_metals}
 \begin{minipage}{6cm}
 \begin{tabular}{cccccc}
  \hline
  Metal & $\left[\frac{X}{\mathrm{H}}\right]_{r_s}$ & Metal & 
$\left[\frac{X}{\mathrm{H}}\right]_{r_s}$ & Metal & 
$\left[\frac{X}{\mathrm{H}}\right]_{r_s}$ \\
  \hline
  C & 8.714 & Fe & 7.996 & Ne & 8.535 \\[2pt]
  Mg & 7.726 & Si & 7.547 & S & 7.457 \\[2pt]
  O & 8.984 & N & 8.175 & Ca & 6.199 \\[2pt]
  \hline
 \end{tabular}
 \begin{tabular}{cccccc}
  & $Z_{4038}/Z_{\sun}$\footnote{Relative to solar metallicities $Z_{\sun}=0.02$} & $Z_{4039}/Z_{\sun}$ & &\\[2pt]
  \hline
   Stellar & 0.992  & 1.053 & &\\[2pt]
   Gaseous & 0.863 & 0.968 & &\\[2pt]
  \hline
  \end{tabular}

 \end{minipage}
\end{table}

To account for the increase of the mass fraction of helium from the
primordial value of $\sim 24\%$ as elements are 
processed during stellar evolution, we apply a metal dependent helium mass fraction.
The helium mass fraction $Y$ has been observed to increase linearly as a
function of the metal mass fraction in stars, $Z=m_\mathrm{met}/m_\mathrm{tot}$,
as $Y \approx 0.24 + 2.4Z$ \citep{2000MNRAS.313...99R, 2007MNRAS.382.1516C}, which is
very close to the relation used in the stellar population synthesis library of
\citet{2003MNRAS.344.1000B} employed in Sec. \ref{section:post_processing}. 
The remaining mass ($\sim 74\%$) in each particle is hydrogen. This yields approximately solar metal mass 
fractions for the galaxies with a slightly higher value for the more compact 
progenitor NGC 4039 (see Table \ref{tab:ic_metals}).

We compare the resulting progenitor galaxies to the observed stellar mass -- gas phase
metallicity ($M_*$--$Z$) relation to verify the realism of the constructed metallicity profiles. 
We use the mean $[\mathrm{O}/\mathrm{H}]$ in the gaseous particles to trace the
metallicity in the initial conditions.
The initial stellar mass of the idealised progenitor galaxies is
$\sim4.7\times 10^{10}$ M$_{\sun}$ and the mean gaseous oxygen abundance is
$\left[\mathrm{O/H}\right]\approx8.8$, placing the progenitors $0.2$ dex below the canonical
observed $M_*$--$Z$ relation
\citep{2004ApJ...613..898T}. However, the type of available data used as a metallicity-indicator 
and the choice of calibration may introduce variations
in the obtained metallicities of up to $0.7$ dex, as pointed out by e.g. \citet{2008ApJ...681.1183K} for
similar SDSS data as used in \citet{2004ApJ...613..898T}. Spectral data is also often
restricted to only the central regions of galaxies due to
 the fibre size of the instruments, which results in a systematical 
overestimation of galactic metallicities \citep{2015MNRAS.454.2381G}.
Thus, given the observational uncertainties, the progenitors are in agreement with the observed $M_*$-$Z$ relations.

Finally, the cosmic formation times of the stellar particles are sorted and 
assigned to the particles according to the metal mass fraction so that the 
least metal-rich particles are given the earliest formation times. The metallicities are given a scatter of $0.2$ dex motivated by the 
the maximum measurement error in \citet{1994A&A...291..757K}, distributed amongst the individual
elements by weighing according to the respective abundances. We set an upper limit for the metal mass fraction of 
$5\%$ for each particle when adding the scatter, to prevent unphysically large 
metallicities for single particles.

\section{Post-processing tools}\label{section:post_processing}

\subsection{Dusty radiative transfer with \skirt}

We produce mock-observations with the panchromatic radiative transfer 
code \skirt\ \citep{2015A&C.....9...20C, 2014A&A...561A..77S}.
The spectral energy distributions (SEDs) produced by each stellar particle are
interpolated within \skirt\ using the metallicity,
age and mass dependent population synthesis models of \citet{2003MNRAS.344.1000B} (BC03 hereafter), assuming a Chabrier initial mass function (IMF)
\citep{2003PASP..115..763C}. \skirt\ uses the SPH data to model the distribution 
of dust, which requires an approximation of the the dust content within the interstellar gas. 
The dust-to-gas ratio in galaxies has been observed to  
correlate with metallicity \citep{1990A&A...236..237I}. Thus, 
in order to obtain the distribution of dust mass in each simulation snapshot processed with \skirt,
we adopt a Milky Way-like dust-to-metals ratio 
of $0.3$ \citet{1998ApJ...501..643D}.
We further assume that only  
gaseous star-forming particles or particles with temperatures below $T<8000$ K are able to 
contain dust, i.e. if a gas particle is too diffuse and too hot we assume that 
the ISM conditions are such that the dust gets destroyed. 

\skirt\ provides multiple
dust models of which we use a Milky Way -like 
implementation with graphite and silicate grains and PAH molecules with a size distribution
adapted from \citet{2004ApJS..152..211Z}. The calorimetric properties of the different grains are adapted from
\citet{2001ApJ...551..807D} and 
the optical properties for the grains are obtained from Bruce Draine's 
website\footnote{http://www.astro.princeton.edu/~draine/dust/dust.diel.html} which summarises results from
various studies (e.g.
\citealt{1984ApJ...285...89D, 1993ApJ...402..441L, 2001ApJ...554..778L}). The spatial
properties of dust are calculated within \skirt\ on an octree grid 
\citep{2013A&A...554A..10S}.

\subsection{Star-forming regions}

When calculating the SED of the simulated
Antennae during the violent phases of the merger, special care must be taken 
with the distribution of the star-forming regions. The young stellar component
embedded in the star-forming regions contributes to the UV-optical part 
of the spectrum, as well as the IR spectrum through dust absorption and 
re-emission. The star formation scheme of \gadget\ is not optimal for tracing 
the distribution of star-forming molecular clouds, since the 
$\sim 10^4$ M$_{\sun}$ stellar particles form stochastically with roughly equal masses.
Observed star-forming regions, on the other hand, 
have a self-similar structure over several orders of 
magnitude in mass, from Jupiter masses to $10^6$ M$_{\sun}$, reflecting the fractal 
nature of interstellar molecular clouds
\citep{1998A&A...331L..65H, 1998A&A...336..697S, 1998A&A...329..249K}. 
For example young stellar clusters in the Antennae have been observed to 
encompass masses from $10^4$ M$_{\sun}$ to $10^6$ M$_{\sun}$ with a mass 
function power-law index of $-2$ 
\citep{1999ApJ...527L..81Z}. 
In modelling the spatial distribution of the youngest stellar population we
follow the implementation of \citet{2016MNRAS.462.1057C} \citep[see also][]{2010MNRAS.403...17J}
with slight modifications
by constructing an additional population of stellar particles to be given as input to \skirt. This particle
population receives distinctive SEDs calculated using the photoionisation code \mappings\ and incorporated 
within \skirt.
Briefly, the SED in these particles represents the distribution of the HII regions and the surrounding 
photo-dissociation regions (PDRs) from which the characteristic spectral signatures of 
star formation are obtained.  For details and the description of the \mappings\ code see
\citet{2008ApJS..176..438G} and references therein.

In order to resolve the radiation field produced by the star-forming regions, 
the distribution of the young stellar component is obtained by grouping and 
redistributing nearby macroparticles into sub-particles with an observationally 
motivated mass distribution.  
First we combine the cold or star-forming gas particles and stellar particles with ages below $10$ 
Myr into gaseous aggregates with masses of $M_\mathrm{tot}\approx10^6$ M$_{\sun}$. 
Given our mass resolution and star formation rates this always results in at least 
a few thousand particles that are eligible for resampling.
Quantities such as metallicity and density
are given by mass weighted averages within each aggregate to regain the properties of the local ISM. 
These aggregates are then resampled into smaller mass particles 
with a mass distribution drawn randomly from the observed mass function of 
interstellar molecular clouds $N\propto M^{-0.8}$ \citep{1998A&A...331L..65H,
1998A&A...329..249K} in the mass range $700$ M$_{\sun}\le M \le 10^6$ M$_{\sun}$ 
and distributed randomly within the volume in which the original particles resided.
In this way the emission from the stochastic particle-based star formation prescription is distributed spatially also into
the gaseous regions where stars could be forming based on the local ISM conditions. As already noted by \citet{2010MNRAS.403...17J},
the mass resolution of the particles with \mappings\ SEDs only have a small effect on the integrated SED of the processed
galaxies. However, we are here interested in the spatial properties of the Antennae merger, for which the redistribution
of star formation is essential.

Next the sub-particles are assigned randomly sampled ages drawn from a cumulative
distribution function assuming a constant SFR for the last $10$ Myr and into the future if necessary.
The assumption of constant SFR is justified as long as the time scale for the rate of change of the 
SFR is longer than the assigned $10$ Myr, which is valid for all snapshots considered here.
The age distribution is calculated simply from the mass of the 
gas aggregate $M_\mathrm{tot}$ and the SFR within the aggregate, given by the total SFR in the gas particles 
used in constructing the aggregate. 
The ages of the sub-particles are then 
used to divide the sub-particles into two categories. The sub-particles which receive ages between $0- 10$
Myr are set as a new particle type, called PDR-particles hereafter. These particles represent the young stellar population in the \gadget\ snapshot
with additional properties required by the \mappings\ SED. Particles which would form in the future,
i.e. the mass in the aggregate not formed during the last $10$ Myr of constant star formation, are returned to the gas population of 
SPH-particles.

The calculation of the SED of the PDR-particles from the tabulated \mappings\ data requires 
the knowledge of the physical state of the molecular cloud in which the young stellar component 
resides. This is achieved by providing \skirt\ with additional parameters for the PDR-particles.
The SFR in an individual PDR-particle is given by the mass $M_\mathrm{PDR}$ 
and the age of the PDR-particle. However, as noted by \citet{2017MNRAS.470..771T}, 
the \mappings\ library assumes that the SFR has been constant for $10$ Myr. Therefore
in order to conserve mass we assume always an age of $10$ Myr when calculating the star formation rates for the PDR-particles.
In order to further preserve the spatial star formation distribution in the snapshots, we also scale the SFR within each aggregate to 
the original SFR value directly derived from the simulation.
The pressure $P$ of the local ISM is obtained from 
the local ISM density $\rho_\mathrm{ISM}$ using an equation of state of the form 
$P\propto \rho_\mathrm{ISM}$ assuming full ionisation. The 
metallicity $Z$ is given by the mass-weighted average metallicity within the combined $10^6$ M$_{\sun}$ gas aggregate.
The compactness $C$ of the HII region is adopted from \citet{2008ApJS..176..438G} and defined as a function of $P$ and $M_\mathrm{PDR}$  
as
\begin{equation}
 \log{C} = 
\frac{3}{5}\log{\left(\frac{M_\mathrm{PDR}}{\mathrm{M}_{\sun}}\right)}+\frac{2}{5}\log{
\left(\frac{P}{k_\mathrm{B } \mathrm{cm^{-3}K}}\right)},
\end{equation}
where $k_B$ is the Boltzmann constant.
The covering fraction of the PDR region $f_\mathrm{PDR}$ 
describes how much of the light escapes directly versus through re-emission. 
The value of the covering factor is set to $f_\mathrm{PDR}=0.1$ following \citet{2016MNRAS.462.1057C},
who validated the value of $f_\mathrm{PDR}$ with a parameter study using similar simulation data to ours.
To obtain the size of the star-forming region, i.e. the size of the smoothing length $h$ of the PDR-particle, 
we employ the cubic spline kernel used in the smoothing of
the properties of the SPH particles within \skirt\ given as 
$M_{\mathrm{cloud}}=(8/\pi) \rho_\mathrm{ISM} h^3$, where $M_{\mathrm{cloud}}$ is the mass of the entire
star-forming region including the surrounding birth cloud.
We invert the
spline kernel to obtain $h$ by assuming that the HII region and the PDR are 
embedded in a gas cloud that is ten times as massive, $M_{\mathrm{cloud}}=10M_{\mathrm{PDR}}$. 
Finally, in order to compensate for the dust in the added gas mass, we also give
\skirt\ an equivalent amount of gas particles with negative mass to be reduced
from the ambient ISM in which PDR-particles are present.

\subsection{Production of mock-images}\label{section:skirt_filters}

\skirt\ is used to produce dust attenuated flux maps at $450$ logarithmically 
spaced wavelengths between $0.02\le \lambda \le 2000$ microns, at different epochs during the 
evolution of the Antennae merger. In all the analysis in this paper we assume a 
distance of $30$ Mpc to the Antennae, which is at the upper range of the observed 
distance estimates, but provides a good match to the morphology of the observed 
Antennae for our chosen initial conditions \citep{2010ApJ...715L..88K}. 

We define the SDSS resolution of $0.396$ arcsec$/$pix as the default resolution of our images both to be
compatible with the \atlas\ survey and to ease the computational load of running
radiative transfer modelling on high resolution \gadget\ snapshots. 
For visualisation purposes, we also process the inner $20$ kpc region of the best match 
Antennae at a Hubble Space Telescope (HST) 
WFPC2-equivalent resolution of $0.1$ arcsec$/$pix and the entire merger including the 
tidal tails at a lower $\sim1$ arcsec$/$pix resolution.
All the images in this paper are oriented such that the upper
and lower galaxies of our best match Antennae
are the observed northern NGC 4038 and the southern NGC 4039, respectively. 

The data cubes obtained from \skirt\ are observed through several filters 
in order to produce mock-images. HST ACS/WFC filters 
F435W, F550M, and F814W 
are used to study the large scale features of the 
best match Antennae, and Johnson-Cousins $U$, $B$, $V$, and $I$ filters 
\citep{1998A&A...333..231B} are used to produce colour composite images of the 
central region. 
The resulting mock-SEDs, mock-RGB composites and derived observables are then compared in 
Sec. \ref{section:overall_appearance} to a wide range of 
observations of the Antennae system such as the HST observations in the Hubble Legacy Archive \citep{1999AJ....118.1551W}.

For the photometric analysis of the merger remnant we employ the SDSS $u$, $g$ and $r$ 
filters using the standard asinh-system of magnitudes\footnote{The asinh-magnitudes are identical to 
the standard logarithmic magnitudes at high flux values (e.g. $\le 20$ mag in the SDSS) but decline less 
rapidly at lower fluxes.} \citep{1999AJ....118.1406L}. 
In order to produce mock-images that are as realistic as possible, we add to the images in each respective
filter an SDSS equivalent background sky component and noise (Pawlik et al., in prep.).
The \skirt\ images in the SDSS bands are used to calculate the colour evolution of the
merger remnant in Sec. \ref{section:photometric_properties} and the produced surface brightness 
profiles are analysed with the fitting tool \galfit\ \citep{2010AJ....139.2097P} in order to
extract characteristic properties directly comparable to observational surveys such as \atlas.

\section{Antennae at the time of best match}

\subsection{Stability of the progenitor galaxies}\label{section:ic_stability}

\begin{figure}
\includegraphics[width=\columnwidth]{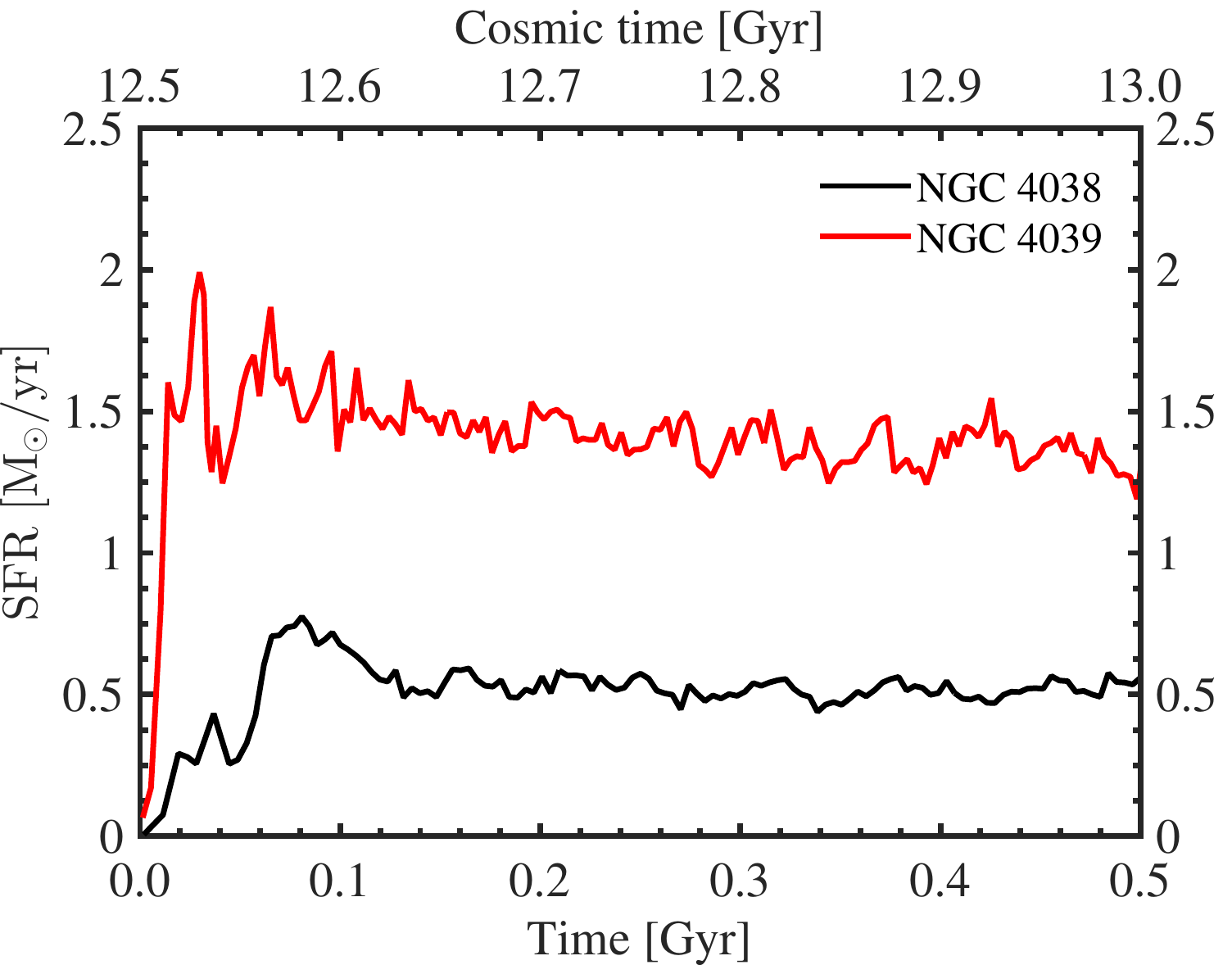}
\caption{The star formation rates of the two isolated progenitor galaxies NGC 4038 (black) and NGC 4039 (red), run separately in isolation for $500$ Myr.
The lower x-axis indicates simulation time and the upper x-axis gives the 
cosmic time.}
\label{fig:ic_sfr}
\end{figure}

We begin by discussing briefly the stability of the initialised progenitor galaxies 
during the orbital approach by analysing the behaviour of the galaxies run in 
isolation for $500$ Myr. 
The total star formation rate (SFR), taken here to be the sum of the subgrid SFR in the
gaseous particles in the simulation, is shown in Fig. \ref{fig:ic_sfr} for each of the progenitors during the $500$ Myr of 
isolated evolution. The progenitors evolve 
as quiescent disc galaxies and the star formation rates quickly settle to 
values of $0.5$ M$_{\sun}/$yr 
for NGC 4038 and $1.5$ M$_{\sun}/$yr for NGC 4039 as was iteratively 
established in setting up the initial conditions. 

\begin{figure}
\includegraphics[width=\columnwidth]{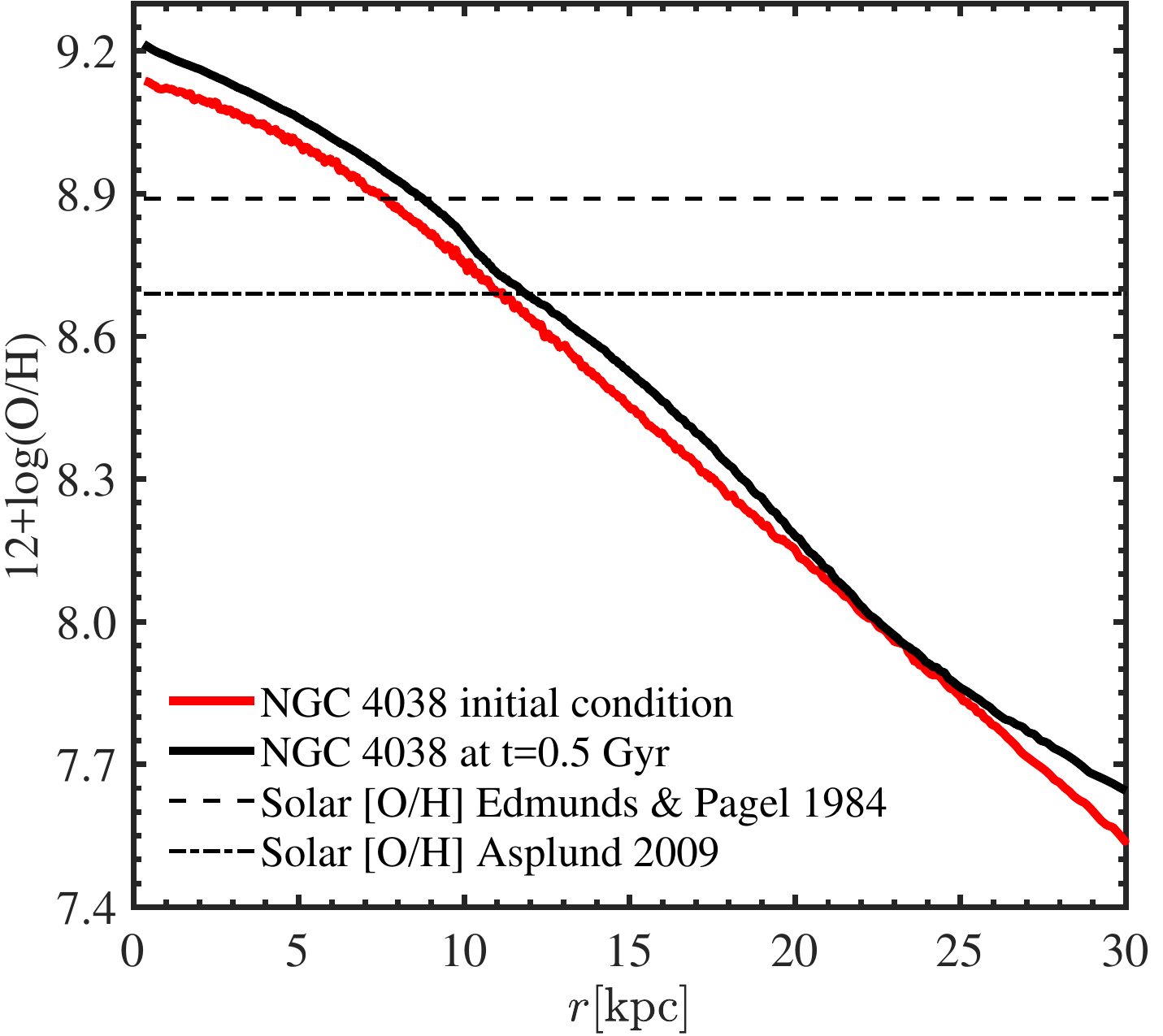}
\caption{The radial oxygen abundance of gas in the isolated progenitor galaxy NGC 4038 run 
in isolation for $500$ Myr. The abundance in the initial conditions is shown in red and
the metallicity after $500$ Myr of evolution is shown in black. The traditional solar abundance
\citep{1984MNRAS.211..507E} is shown as a dashed line, together with the updated value of 
\citet{2009ARA&A..47..481A} shown as a dot-dashed line.}
\label{fig:ic_metals}
\end{figure}

The initial radial oxygen abundance of the more 
extended progenitor galaxy NGC 4038 is shown in Fig. \ref{fig:ic_metals}, 
compared with the abundance after 
$500$ Myr of isolated evolution. 
For the NGC 4039 progenitor, the behaviour is very similar and therefore not shown separately.
The radial abundance gradient is initially slightly flatter in the central region due to the maximum metal  
mass fraction of $5\%$ per particle defined in the initial conditions. 
The metallicity gradients of the gaseous and stellar 
discs are retained as the progenitor galaxy evolves in isolation. As a result, the gaseous oxygen abundance of NGC 4038 progenitor 
is enriched by less than $0.1$ dex during the $500$ Myr timespan.

\begin{figure}
\includegraphics[width=\columnwidth]{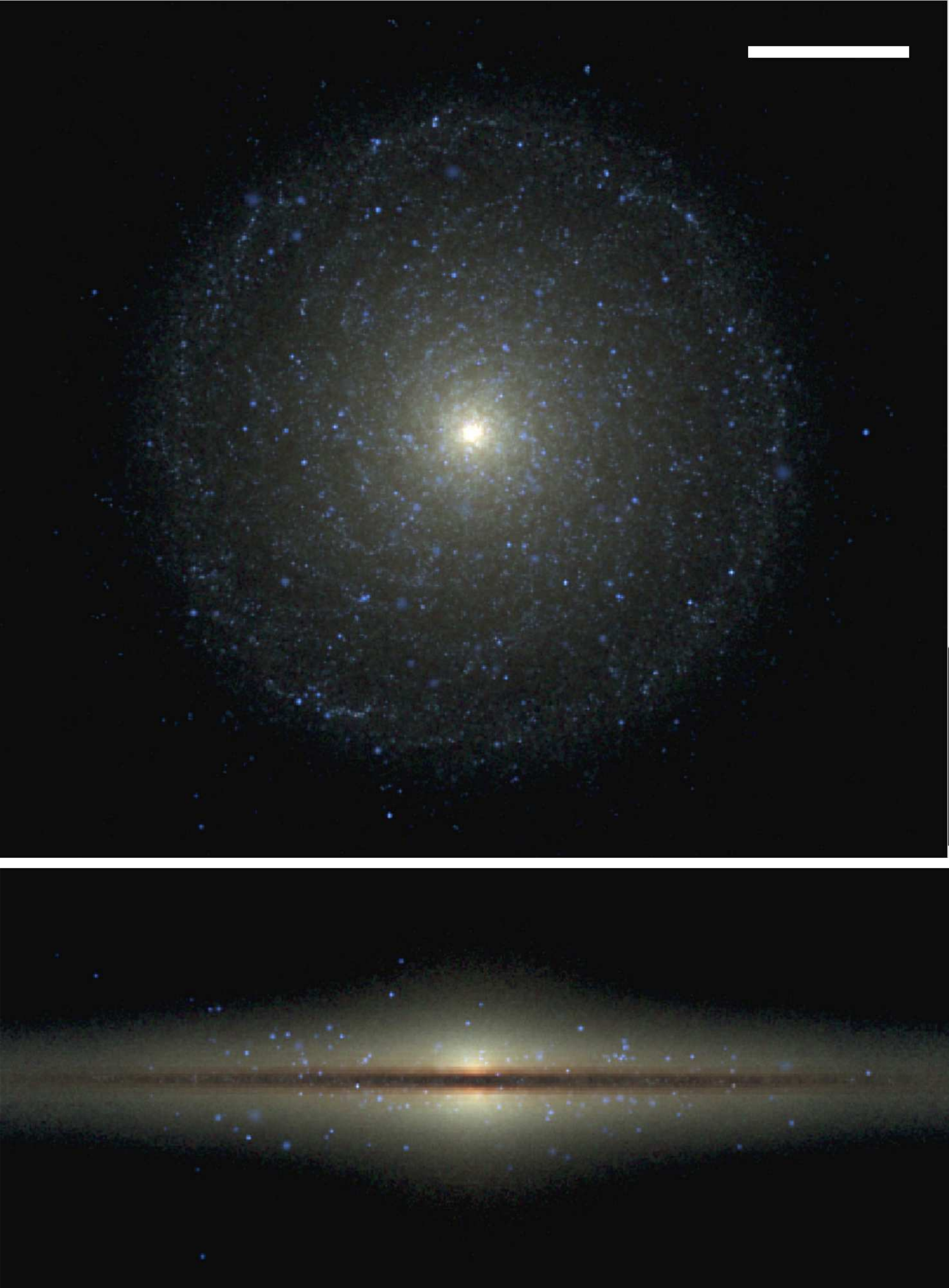}
\caption{Mock face-on (top) and edge-on (bottom) images of the 
NGC4038 progenitor at a typical SDSS resolution of $0.396$ arcsec$/$pix through Johnson $U+B$ (blue), 
$V$ (green) and $I$ (red) filters after $500$ Myr of isolated evolution. 
The images span $30$ kpc horizontally and the bar indicates a scale of $5$ kpc. The images are produced using \skirt\ (see Sec. 
\ref{section:post_processing}) including the resampling of the young stellar particles.}
\label{fig:NGC4038_progenitor}
\end{figure}

Mock multi-colour face-on and edge-on images of the NGC 4038 progenitor in the Johnson-Cousins $U+B$, $V$ and 
$I$ bands, produced using the methods described in Sec. \ref{section:post_processing}, are shown 
in Fig. \ref{fig:NGC4038_progenitor}.
As is evident from Figures \ref{fig:ic_sfr}--\ref{fig:NGC4038_progenitor}, 
the progenitor discs remain stable against excessive fragmentation during the $500$ Myr simulation timespan. 
No large scale disc instabilities are seen, which would be able to drive massive gaseous outflows
or cause unpredictable amounts of star formation.

\subsection{Overall appearance}\label{section:overall_appearance}

\begin{figure*}
\includegraphics[width=0.9\textwidth]{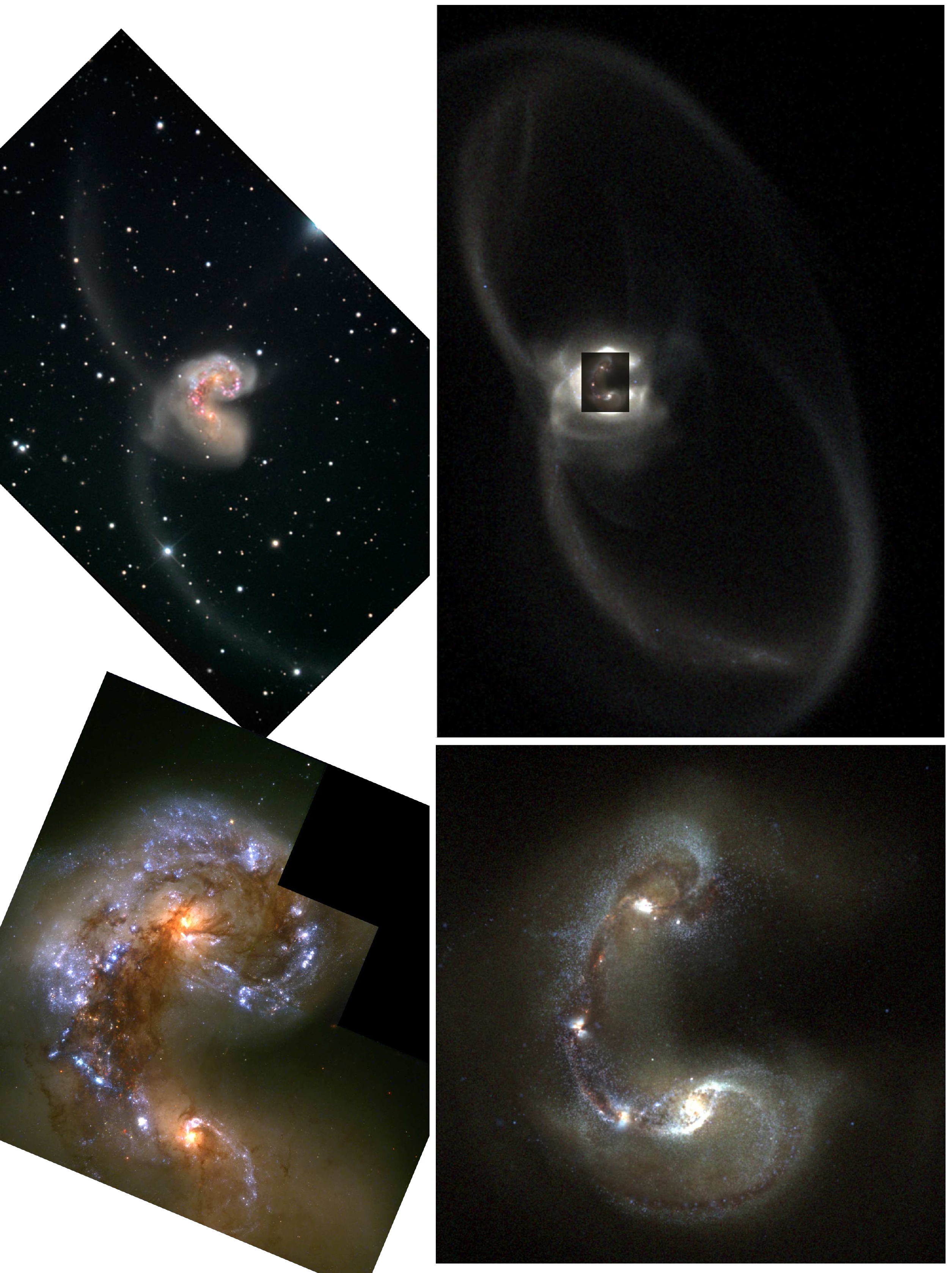}
\caption{Left column: a colour composite from the Hubble Legacy Archive in ACS/WFC F435W (blue), 
F550M (green), F658N ($H_\alpha$+[N II]) (pink) and 
F814W (red) filters (top), and a HST composite of the inner region 
converted to Johnson-Cousins $U+B$ (blue), $V$ (green) 
and $I$ (red) bands from \citet{1999AJ....118.1551W} (bottom). Right column: simulated 
colour composite images of the entire Antennae 
system and its inner region obtained through the same 
respective filters (omitting the $H_\alpha$-band). The resolution of the mock-image  
is $1.03$ arcsec$/$pix for the large scale image (top)
and a WFPC2 equivalent resolution of $0.1$ arcsec$/$pix for
the inner region (bottom). The tails have been highlighted using a logarithmic stretch, whereas the inlet and the
zoomed image have a square root stretch.
The intricate details in the simulated images are best viewed on a computer screen.}
\label{fig:mock_Antennae}
\end{figure*}

The best match of our simulated Antennae 
model is produced $530$ Myr after the first and $41$ Myr after the second 
passages, $80$ Myr before the final coalescence of the nuclei.
We show the post-processed mock-images of the best match Antennae
in Fig. \ref{fig:mock_Antennae}, along with two
HST composite images: an image showing the large scale features 
obtained from the Hubble Legacy Archive and a more detailed image showing
the inner region of the merger produced by 
\citet{1999AJ....118.1551W}. The mock-images have been produced using the same ACS/WFC and Johnson 
filters as used in the respective observed images, as described in Sec. \ref{section:skirt_filters}.

The most striking features of the observed Antennae,
such as the tidal tails and the overlap region, are clearly visible in the 
mock HST images. The dust obscuration in the overlap region and the 
tidal arcs around the still separate nuclei follow the presence of star-forming gas.  
Blue light from the star-forming regions is present
in similar regions and with the same clumpy structure in both the simulated and observed 
images. Yellow, dispersed light from the older stellar population is seen as 
a more extended, ambient background compared to the narrow dust lanes and
blue young stellar populations.

HST observations at high resolution allow us to examine 
the central region in more detail and interpret the origin of the various
characteristic regions using the simulation and the mock-observations as references.
A large number of $500$--$600$ Myr old stellar populations have been 
observed in the Antennae, indicating the time of the first passage
\citep{1999AJ....118.1551W}. 
Young to intermediate age clusters with ages below $30$ Myr are found in the loop 
structures around NGC 4038 and in the overlap region indicating the time of the second 
passage \citep{2009ApJ...701..607B}, and the very youngest clusters are found in the 
overlap region \citep{2009ApJ...699.1982B}.  Interpreting the 
evolution of the simulated Antennae, we can identify the observed 
overlap region as the tidal bridge formed during the second passage of the 
galaxies. The intermediate age stellar clusters in the loop structures are less 
prominent tidal tail -like structures formed during the second passage, whereas the observed older 
population of clusters matches with our timing for the first passage and the 
formation time of the extended tidal tails.

\subsection{Spectral energy distribution}

\begin{figure*}
\includegraphics[width=\textwidth]{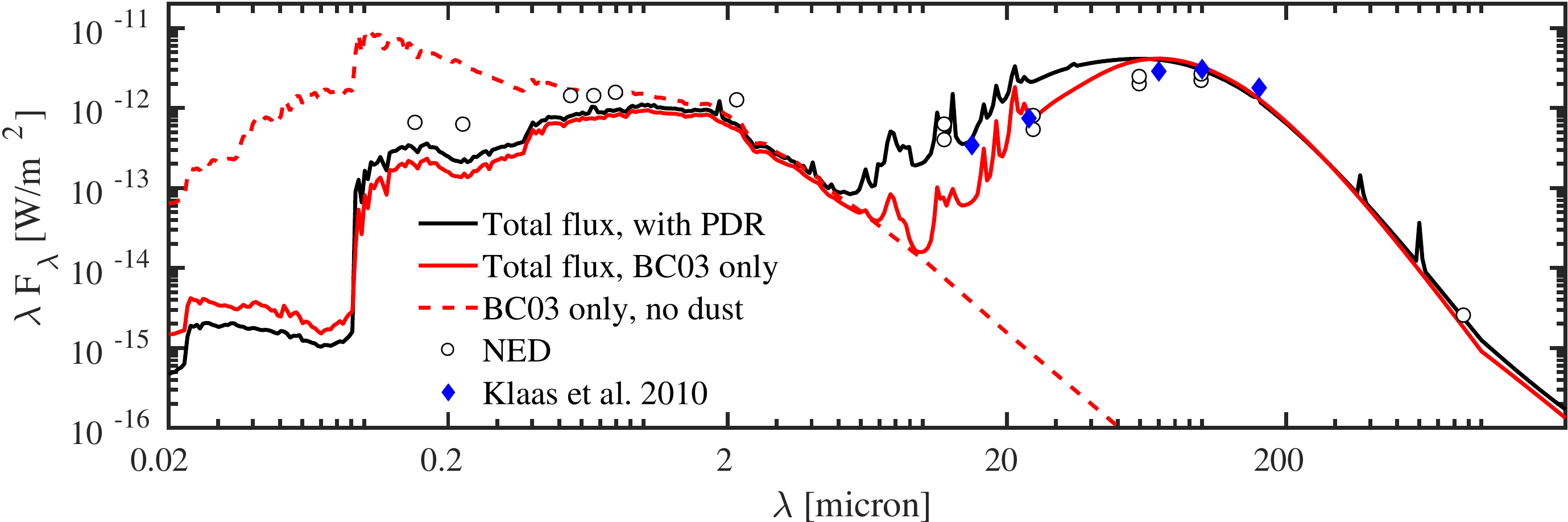}
\caption{The mock spectral energy distribution of the simulated Antennae with (solid black) and without (solid red)
the resampling of the young stellar component, produced with \skirt. Both of these models include dust, whereas
the dashed red curve shows the BC03 flux without dust attenuation. 
Observed fluxes in different wavebands obtained from NED (open markers) and \citet{2010A&A...518L..44K} (blue markers) are shown for reference. 
The error bars in the observations
are smaller than the marker size. Multiple datapoints at the same wavelength indicate the lower and upper limits of the observed
values.}
\label{fig:Antennae_SED}
\end{figure*}

The spectral energy distribution of the main body of the simulated 
Antennae is shown in Fig. \ref{fig:Antennae_SED} along with the observed values obtained from 
NED\footnote{The NASA/IPAC Extragalactic
Database (NED) is operated by the Jet Propulsion Laboratory, California Institute of Technology,
under contract with the National Aeronautics and Space Administration.} and \citet{2010A&A...518L..44K}. In Fig. \ref{fig:Antennae_SED}
we show the stellar BC03 flux with (solid red line) and without dust attenuation (dashed red line). If dust attenuation is omitted, the simulated UV-flux 
would overproduce the observed values by over an order of magnitude. Including dust reprocesses the radiation shifting a significant amount of the radiation 
to longer wavelengths, resulting now in turn in an underestimation of the UV-optical flux by a factor of $2$--$3$.
In Fig. \ref{fig:Antennae_SED} we also show the SED calculated both with (solid black line) and without the resampling of the young
stellar component (solid red line) in order to show the effect of adding the \mappings\ SEDs.
Direct radiation from the new population of resampled star formation regions contributes to the flux in the UV-optical
part of the spectrum, increasing the flux by up to 50\% towards the observed values. 
However, even adding the \mappings\ SED will result in an UV-optical flux which is underestimated by a factor of $\sim 2$ with respect to the observations. 
Dust included around the young star-forming regions shifts again some of the emission from the young stars to infrared wavelengths, 
which brings additional contribution to the infrared part of the spectrum. As a result, the final infrared flux matches the 
observed flux in the mid-infrared at $\lambda=12$ $\mu$m and at long wavelengths of $\lambda>100$ $\mu$m. However, the simulated 
flux is overestimated by up to a factor of $3$ in the intermediate infrared range of $20 < \lambda < 100$ $\mu$m.
We have adopted a distance of $30$ Mpc following \citet{2010ApJ...715L..88K}, 
however modifying this closer to the observationally favoured value of $20$ Mpc would
increase the flux by a factor of $\sim 2$, increasing the optical flux close to the
observed values, but would result in an even more severe overprediction of the IR flux.

We note that \citet{2013MNRAS.434..696K} also modelled the far-infrared spectrum of their simulated Antennae
galaxies using a dust model similar to ours. 
They were able to produce SED shapes similar to the Herschel observations by assuming a universal gas-to-dust ratio
two times smaller than the canonical value of $124$:$1$, yet still underproduced the level
of far-infrared emission by at least $\sim 20$--$30\%$. Our assumption for a dust-to-metals ratio of $0.3$ 
yields values ranging from $68$:$1$ to $472$:$1$ for our gas-to-dust ratio within the central $30$ kpc. This is in a similar range as
used in previous studies, but here rather than setting a uniform gas-to-dust ratio for all gaseous particles, we have a distribution
of values based on the evolved metallicity of the simulation particles. 

The B-band luminosity of the post-processed galaxy is $L_{\rm B}=4.16 \times 10^{10}$ 
L$_{\sun}$. Given the large uncertainty in the distance estimate,
this is consistent with the 
observed value of $L_{\rm B}=2.9\times10^{10}$ L$_{\sun}$ \citep{2001AJ....122.2969H}
measured assuming a distance of $19.2$ Mpc. 
The derived $V$-band magnitudes of the simulated Antennae at $20$ and 
$30$ Mpc distance are $10.2$ and $11.1$ mag, respectively. This is consistent
with the observed $V$-band magnitude of $10.37$  \citep{1988cvip.book.....D}.
All the observables derived from the simulations have been summarised 
in Table \ref{tab:Antennae_observations} along with the corresponding observed values.

\subsection{Star formation rate}\label{section:sfr_observations_of_Antennae}

\begin{figure}
\includegraphics[width=\columnwidth]{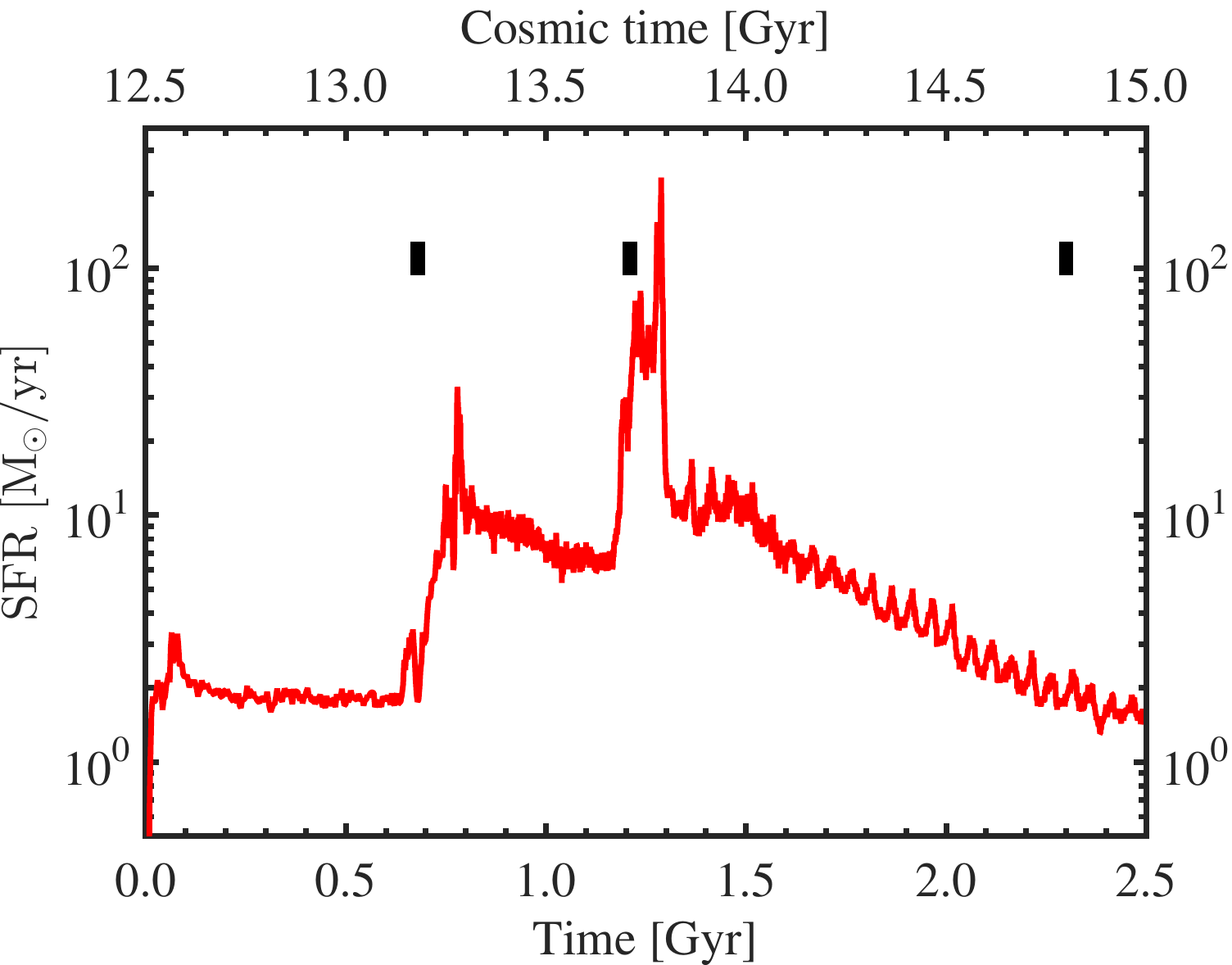}
\caption{The star formation rate during the first $2.5$ Gyr of the simulation, taken
as the sum of the subgrid star formation rates in the gaseous particles, averaged over 500 
time steps. The lower x-axis indicates simulation time and the upper x-axis gives the 
cosmic time. The vertical bars indicate the time of the first 
pericentric passage, the time of the best match to observations ($\sim13.7$ 
Gyr) and the remnant age of $1$ Gyr used for the majority of the analysis related to the merger remnant.}
\label{fig:Antennae_sfr}
\end{figure}

The total SFR during the simulation is shown in Fig. \ref{fig:Antennae_sfr}, taken again as the sum of the gaseous subgrid SFR in the simulation and
smoothed over 500 time steps which corresponds to approximately $0.2$--$3$ Myr.
The vertical bars in Fig. \ref{fig:Antennae_sfr} indicate the first passage $\sim 530$ Myr ago, the present epoch at 
$\sim13.7$ Gyr, and when the remnant reaches an age of $1$ Gyr, respectively, assuming a Hubble constant of $H_{\rm{0}}=71$ km s$^{-1}$ Mpc$^{-1}$. 
The peak SFR during and immediately after the coalescence some $80$ Myr in the future
reaches up to $\sim 200$ M$_{\sun}/$yr. During the following few Myr after the coalescence, the total SFR
declines gradually and decreases to a value of a few 
solar masses per year after most of the gas ejected during the encounter 
falls back to the remnant. Some star
formation is present in the remnant even at later age, situated mainly within the inner $1$--$2$ kpc
in a nuclear disc that forms after the final coalescence.

\begin{figure}\includegraphics[width=\columnwidth]{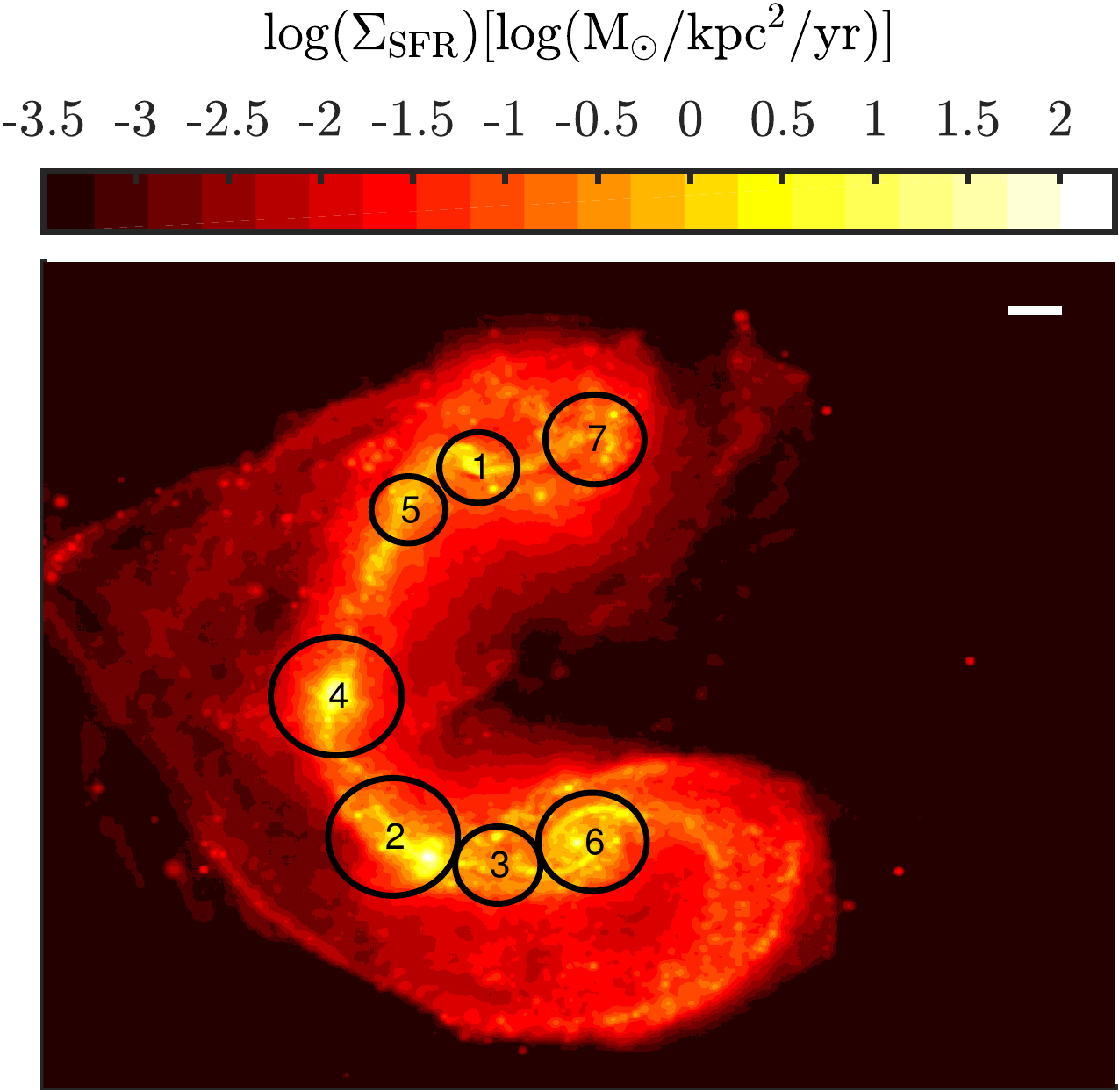}
\caption{The spatial star formation rate density of the best match Antennae, as integrated 
on pixel-by-pixel basis from the IR-luminosity (see Sec. \ref{section:sfr_observations_of_Antennae}). 
The circles encompass the IR-bright knots in the galactic nuclei (1 and 6), the overlap region (2-5) and the northern stellar arc (7)
and they indicate the apertures within which the SFRs are integrated. 
The image spans $20$ kpc, the white bar in the top-right corner indicates a scale of $1$ kpc, and
the image resolution is the HST ACS/WFPC2 equivalent of $0.1$ arcsec$/$pix.}
\label{fig:Antennae_spatial_sfr}
\end{figure}

Following \citet{2010ApJ...715L..88K,2013MNRAS.434..696K}, we also analyse the 
spatial distribution of star formation in the central region of the simulated merger. 
Observations have shown a peculiar feature, where the lower limit SFR 
(sum over the brightest IR knots) in the overlap region exceeds the SFR of both of the 
nuclei \citep{2009ApJ...699.1982B, 2010A&A...518L..44K}. This feature is also reproduced in 
our simulation, as can be seen in Fig. \ref{fig:Antennae_spatial_sfr}, where 
the spatial SFR density of the best match Antennae is shown. To make a connection to the observations, 
the SFR density in Fig. \ref{fig:Antennae_spatial_sfr} has been calculated from the IR-luminosity
integrated from the \skirt\ SED on a pixel-by-pixel basis using the relation 
\begin{equation}\label{eq:kennicutt}
 \mathrm{SFR}=1.15\times10^{-10} \left(\frac{L_{\mathrm{IR}}}{\mathrm{L}_{\sun}}\right) \mathrm{M}_{\sun}/\mathrm{yr},
\end{equation}
where $L_{\rm IR}$ is the luminosity in the wavelength range $8-1000$ micron. The definition in Eq. \ref{eq:kennicutt} is the
traditional Kennicutt-relation from \citet{1998ApJ...498..541K} divided by $1.5$
to accommodate for the use of the Chabrier IMF \citep{2007ApJS..173..267S, 2007ApJS..173..315S}.
Following \citet{2010A&A...518L..44K} we have chosen $7$ IR-bright regions (indicated with 
numbers $1$--$7$ in
Fig. \ref{fig:Antennae_spatial_sfr}) with circular apertures of areas similar to the ones specified in
Table 3 of \citet{2010A&A...518L..44K}. The integrated star formation rates are summarised in Table \ref{tab:Antennae_observations}. 

The simulated star formation rates in the nuclei (regions $1$ and $6$) are SFR$_\mathrm{4038}\approx1.0$ M$_{\sun}/$yr and 
SFR$_\mathrm{4039}\approx2.5$ M$_{\sun}/$yr, whereas the total SFR in the IR-knots in the
overlap region (regions $2$--$5$) is SFR$_\mathrm{overlap}\approx15.8$ M$_{\sun}/$yr. The corresponding observed values
in \citet{2010A&A...518L..44K} are SFR$_\mathrm{4038}\approx1.17$ M$_{\sun}/$yr, 
SFR$_\mathrm{4039}\approx0.5$ M$_{\sun}/$yr and SFR$_\mathrm{overlap}\approx4.84$ M$_{\sun}/$yr.
On the other hand \citet{2009ApJ...699.1982B} derive observed values for the SFR that are   
approximately 50\% lower than the SFR values of \citet{2010A&A...518L..44K} 
in the nuclei of the galaxies while they obtain for the total overlap region a SFR of $5.4$ M$_{\sun}/$yr. 
The star formation rate in the north-eastern region of NGC 4038 (region $7$) is measured to be $0.55$ M$_{\sun}/$yr
in our simulation and it is comparable to the observed value of $0.5$ M$_{\sun}/$yr.
The study by \citet{2013MNRAS.434..696K} also derived star formation rates in similar regions of their simulated Antennae but only
by directly taking the sum of the subgrid SFRs of the simulated gas particles. Varying the efficiency of their stellar feedback
model \citet{2013MNRAS.434..696K} found simulated SFR values of $10.88$ M$_{\sun}/$yr $\le$ SFR$_\mathrm{overlap}\le26.09$ M$_{\sun}/$yr for 
the overlap region, and $0.49$ M$_{\sun}/$yr $\le$ SFR$_\mathrm{4038}\le0.92$ M$_{\sun}/$yr, $0.44$ M$_{\sun}/$yr $\le$ SFR$_\mathrm{4039}\le2.02$ M$_{\sun}/$yr
for the nuclear regions of NGC 4038 and and NGC 4039, respectively.

Some of the regions in our simulation show therefore star formation on similar scale to the observations, whereas e.g. the star-forming knots 
in the simulated overlap region overestimate the SFR compared to observations.
The total SFR of the entire Antennae has been observed to span values between $5$--$22.2$ 
M$_{\sun}/$yr \citep{1990ApJ...349..492S, 2001ApJ...561..727Z, 2009ApJ...699.1982B, 2010A&A...518L..44K}. 
The lowest values of SFR are measured as the sum of the SFR in the observed IR-knots, while 
e.g. \citet{2010A&A...518L..44K} observe an SFR twice as high outside the few brightest IR regions.
Our estimate of the total SFR is $\sim 23.6$ M$_{\sun}/$yr,
which is only somewhat above the upper limit of observed values. The star formation is however concentrated
in rather tight knots within the brightest regions, whereas in the observed Antennae the star formation appears 
to be somewhat more diffuse.

The most important feature observed in the Antennae is the overlap starburst where the observed lower
limit value for the ratio of star formation in the overlap region relative to the nuclei 
is $(\mathrm{SFR}_{\mathrm{overlap}}/\mathrm{SFR}_{\mathrm{nuclei}})_{\mathrm{obs}}=2.9$--$5.4$ 
\citep{2009ApJ...699.1982B, 2010A&A...518L..44K}. This is also reproduced 
in our simulation for which we find
$(\mathrm{SFR}_{\mathrm{overlap}}/\mathrm{SFR}_{\mathrm{nuclei}})_{\mathrm{sim}}\approx4.5$. In comparison,
\citet{2013MNRAS.434..696K} 
obtained $9.1\le(\mathrm{SFR}_{\mathrm{overlap}}/\mathrm{SFR}_{\mathrm{nuclei}})_{\mathrm{obs}}\le11.1$ for the ratio of SFR between 
the overlap region and the nuclei, which is larger than the observed ratio by a factor of $\gtrsim 2$--$3$.

\begin{table*}
 \caption{Derived observational properties of the best match snapshot processed through 
 \skirt, compared to the observed values. The values are derived assuming a distance of 
 $30$ Mpc where necessary and values in parentheses give the derived value at a distance of $20$ Mpc. 
 SFR values were calculated from the integrated IR-luminosity as described in Sec. \ref{section:sfr_observations_of_Antennae},
 and the value for the overlap region is a lower limit from the combined SFR in the four IR-bright knots in
 Fig. \ref{fig:Antennae_spatial_sfr}.
 }
 \label{tab:Antennae_observations}
 \begin{minipage}{26cm}
 \begin{tabular}{lllll}
 \hline
   & Derived value & Observed value & References & Comments \\
  \hline
    $m_\mathrm{V}$  & $11.1$ ($10.2$) & $10.37$ & \citet{1988cvip.book.....D} &  \\[2pt]
    $L_\mathrm{B}$  & $4.16\times10^{10}$ L$_{\sun}$ ($1.9\times10^{10}$ L$_{\sun}$) & $2.9\times10^{10}$ L$_{\sun}$ & 
    {\citet{1991rc3..book.....D}} \\[2pt]
    SFR$_\mathrm{total}$  & $23.6$ M$_{\sun}/$yr & $22.2$ M$_{\sun}/$yr & & \\[2pt]
    SFR$_\mathrm{overlap}$ & $15.8$ M$_{\sun}/$yr & $4.84$ M$_{\sun}/$yr & 
    {\citet{2010A&A...518L..44K}} &  Regions $2$--$5$ in Fig. \ref{fig:Antennae_spatial_sfr} \\[2pt]
    SFR$_\mathrm{4038}$  & $1.0$ M$_{\sun}/$yr & $1.17$ M$_{\sun}/$yr & 
    {\citet{2010A&A...518L..44K}} & Region $1$ in Fig. \ref{fig:Antennae_spatial_sfr} \\[2pt]
    SFR$_\mathrm{4039}$  & $2.5$ M$_{\sun}/$yr & $0.5$ M$_{\sun}/$yr & 
    {\citet{2010A&A...518L..44K}} & Region $6$ in Fig. \ref{fig:Antennae_spatial_sfr} \\[2pt]
    SFR$_\mathrm{arm}$  & $0.55$ M$_{\sun}/$yr & $0.5$ M$_{\sun}/$yr & 
    {\citet{2010A&A...518L..44K}} & Region $7$ in Fig. \ref{fig:Antennae_spatial_sfr} \\[2pt]
    SFR$_{\mathrm{overlap}}/\mathrm{SFR}_{\mathrm{nuclei}}$  & $4.5$ & $ 2.9 $ & 
    {\citet{2010A&A...518L..44K}} &  \\[2pt]
  \hline
 \end{tabular}
 \end{minipage}
\end{table*}

\subsection{Metallicity}\label{section:antennae_metallicity}

High-resolution spectroscopy of stellar clusters within the Antennae galaxies 
has enabled studies of the spatial metallicity distribution in the ongoing merger. In general, metallicities ranging from slightly 
sub-solar to super-solar ($0.9<Z/Z_{{\sun}}<1.3$) have been found for young to intermediate age ($ < 6\times 10^8$ yr)
stellar clusters with typical uncertainties of $0.1$--$0.2$ dex \citep{2009ApJ...701..607B, 2015ApJ...812..160L}. 
As noted by \citet{2008ApJ...681.1183K} and verified for the stellar clusters in the Antennae by
\citet{2015ApJ...812..160L}, additional variations arise from the specific metallicity-indicator and 
calibration methods used, which may result in offsets up to $0.7$ dex for the measured metallicities. 

\begin{figure}\includegraphics[width=\columnwidth]{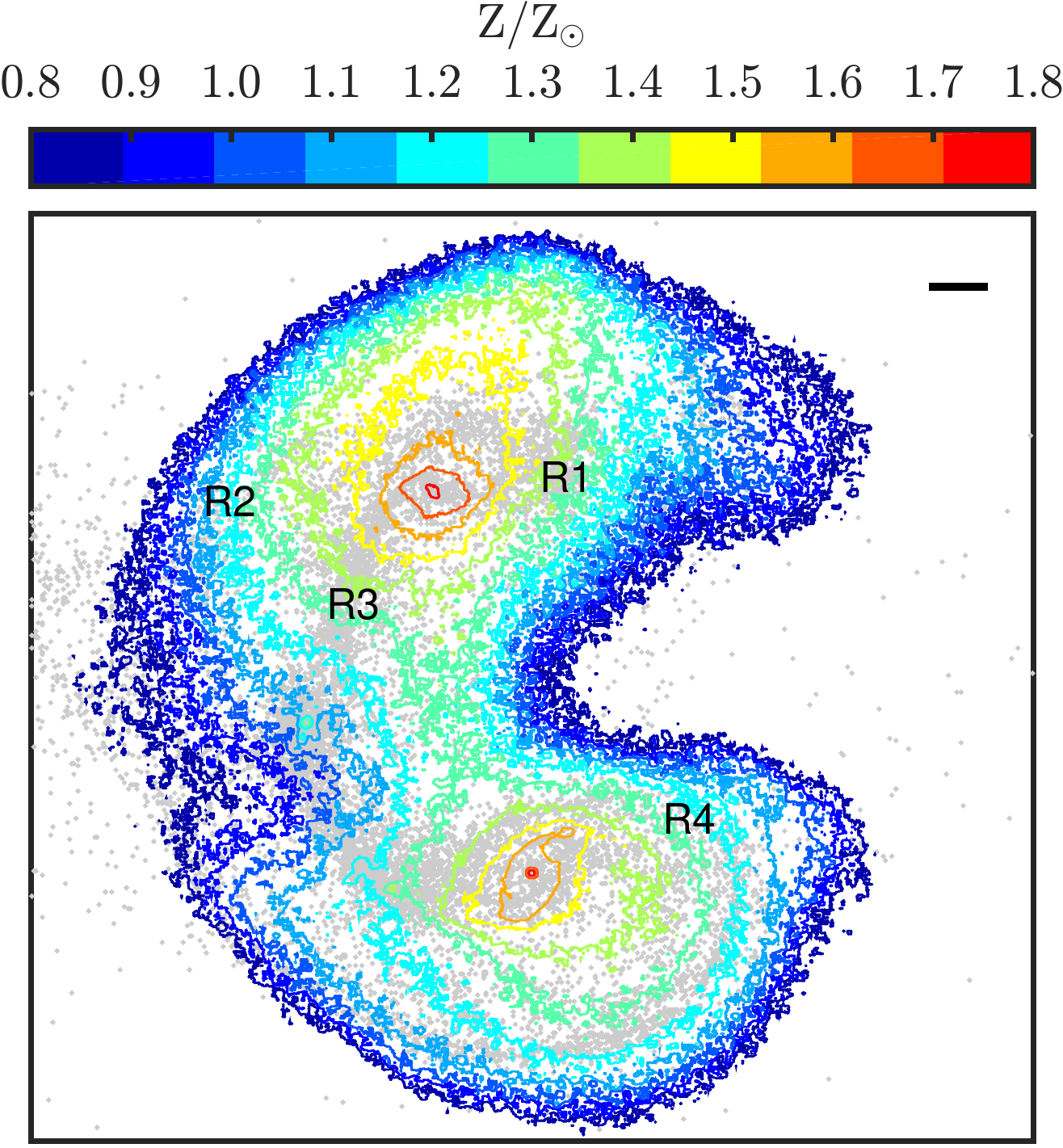}
\caption{The contours of the pixel-by-pixel mass-weighted mean stellar metallicity of the best match Antennae
in solar units for $Z_{\sun}=0.02$. Underlaid grey points show the positions of every 5th randomly chosen
gas particle. Labels R1--R4 show the approximate positions of the
characteristic regions where stellar clusters have been observed (see text for details).
The image spans $20$ kpc and the black bar in the top-right corner indicates a scale of $1$ kpc.}
\label{fig:Antennae_spatial_metals}
\end{figure}

Given the uncertainties in the observed values, we calculate here the metallicity based only on the
simulated properties of the particles and leave a more detailed examination 
using more observationally motivated metallicity proxies for a future study. The mass-weighted
spatial metallicity of stellar particles within the central $10$ kpc of the simulated Antennae is shown in Fig.
\ref{fig:Antennae_spatial_metals} in solar units (assuming $Z_{\sun}=0.02$). The image is smoothed to a resolution of $100$ pc,
corresponding to the spatial resolution in the \citet{2009ApJ...701..607B} study.
Positions of every 5th randomly selected gas particle are underlaid in Fig. \ref{fig:Antennae_spatial_metals}
to aide the comparison of the matter distribution of our best match Antennae and the observed regions.

Most of the structures within the central $10$ kpc show super-solar metallicities, with the nuclear
regions reaching metallicities nearly twice the solar value. 
In addition to the nuclear regions, four regions where stellar clusters have been observed 
have been identified in our best match image, as indicated in Fig. \ref{fig:Antennae_spatial_metals} with 
labels R1-R4. 
Region 1 represents the arc around the nucleus of NGC 4038, region 2 represents the radial region outside the brightest 
optical emission at a distance of $4$--$5$ kpc from the NGC 4038 nucleus, region 3 is the northern part of the
overlap region at $3$ kpc distance from the NGC 4038 nucleus, and region 4 represents the arc around the nucleus of NGC 4039. The observed metallicity 
ranges with maximum errors for these regions are 
$Z_\mathrm{4038, obs}=1.3\pm0.2$ Z$_{\sun}$, $Z_\mathrm{4039, obs}=1.1\pm0.2$ Z$_{\sun}$, $Z_\mathrm{R1, obs}=1.2$--$1.3\pm0.2$ Z$_{\sun}$ 
($1.0$--$1.3 \substack{+0.19 \\ -0.23}$ Z$_{\sun}$),
$Z_\mathrm{R2, obs}=1.0$--$1.1\pm0.4$ Z$_{\sun}$, $Z_\mathrm{R3, obs}=1.1$--$1.3\pm0.2$ Z$_{\sun}$ ($1.2 \substack{+0.21 \\ -0.25}$ Z$_{\sun}$)
 and $Z_\mathrm{R4, obs}=0.9$--$1.1\pm0.1$ Z$_{\sun}$, respectively,
from \citet{2009ApJ...701..607B} and in brackets from \citet{2015ApJ...812..160L}. 

The range of values for the resulting simulated metallicities within the nuclei and regions R1--R4 are studied here within $1$ kpc circular apertures.
The nuclei, where the star formation is strongly concentrated, show enhanced
metallicities with $Z_\mathrm{4038, sim}=1.5$--$1.8$ Z$_{\sun}$ and $Z_\mathrm{4039, sim}=1.4$--$1.8$ Z$_{\sun}$.
The derived range in the arc region at R1 is $Z_\mathrm{R1, sim}=1.1$--$1.5$ Z$_{\sun}$. The outer regions at
R2 and likewise at the opposite side of the NGC 4038 nucleus have a metallicity in the range $Z_\mathrm{R2, sim}=0.8$--$1.4$ Z$_{\sun}$. The overlap region
includes a wide range of values from $0.9$ Z$_{\sun}$ to $1.5$ Z$_{\sun}$, with R3 at $Z_\mathrm{R3, sim}=1.2$--$1.5$ Z$_{\sun}$. 
The arc around the nuclear region of NGC 4039
is somewhat larger than the corresponding observed area (see Fig. \ref{fig:mock_Antennae}), but the region around R4 and along the gaseous arc shows metallicities 
in the range of $Z_\mathrm{R4, sim}=1.0$--$1.4$ Z$_{\sun}$.
The off-nuclear regions result therefore in metallicities in broad agreement with the observed range,
while the nuclear metallicities are enhanced by $0.5$--$0.7$ Z$_{\sun}$ compared to the observed nuclei.
A more observationally consistent method would require detailed spectral mock data and a stellar cluster selection based on age, 
as the observed super stellar clusters are limited to ages of $t <6.3\times10^8$ yr.
Most importantly, the agreement on the large spatial scale adds to the credibility of our initial conditions.
We thus conclude that the assumption of Milky Way -like progenitor metallicity acts as a reasonable starting point for further studies.

\section{Intrinsic structure of the merger remnant}\label{section:intrinsic_structure}

\subsection{Intrinsic shape and the density profile}\label{section:intrinsic_shape}

The merger remnant is predominantly studied $\sim1$ Gyr after the final coalescence of the nuclei at a simulation 
time of $t=2.3$ Gyr. During the merger, the stellar mass has grown by $1.02\times10^{10}$ 
M$_{\sun}$, corresponding to an $\sim11$\% increase over the initial combined stellar mass, whereas only 
$4.05\times10^{9}$ M$_{\sun}$ of the gas mass, corresponding to $\sim25$\% of the initial value remains 
in a gaseous phase. The central 30 kpc region of the remnant consists 
of $1.42\times10^{9}$ M$_{\sun}$ gas and $8.82\times10^{10}$ M$_{\sun}$ stellar 
mass, which amounts to $\sim35$\% and $\sim85$\% of the total gaseous and 
stellar mass in the simulation volume. The remainder of the baryonic mass resides in the tidal tails and 
the gaseous rings around the merger remnant. 

We start by studying the intrinsic three dimensional structure of the merger remnant. 
A discussion of the projected apparent shape, taking into account observable effects
such as dust attenuation, is deferred to Sec. \ref{section:photometric_properties}. 
Here we use the unweighted ellipsoidal shell method, i.e. the S1 method 
of \citet{2011ApJS..197...30Z} to define the intrinsic shape of the stellar component of the remnant.
In this method the axial ratios $b/a$ and $c/a$ are calculated iteratively
in thin ellipsoidal shells to obtain the local shape as a function of radius.
The calculation reveals a relatively spherical central structure with axis ratios $b/a$ and $c/a$
predominantly in the range $0.8$--$0.9$ within the inner $10$ kpc. The axis ratio profiles peak within $\sim0.5$ kpc
with maximum values of $b/a=0.93$ and $c/a=0.91$ and decrease gradually to values of $b/a=0.88$ and $c/a=0.8$ towards
the outer radii, thus showing a spherical inner structure which transitions into a slightly more elongated shape in the outer parts.

\begin{figure}
\includegraphics[width=\columnwidth]{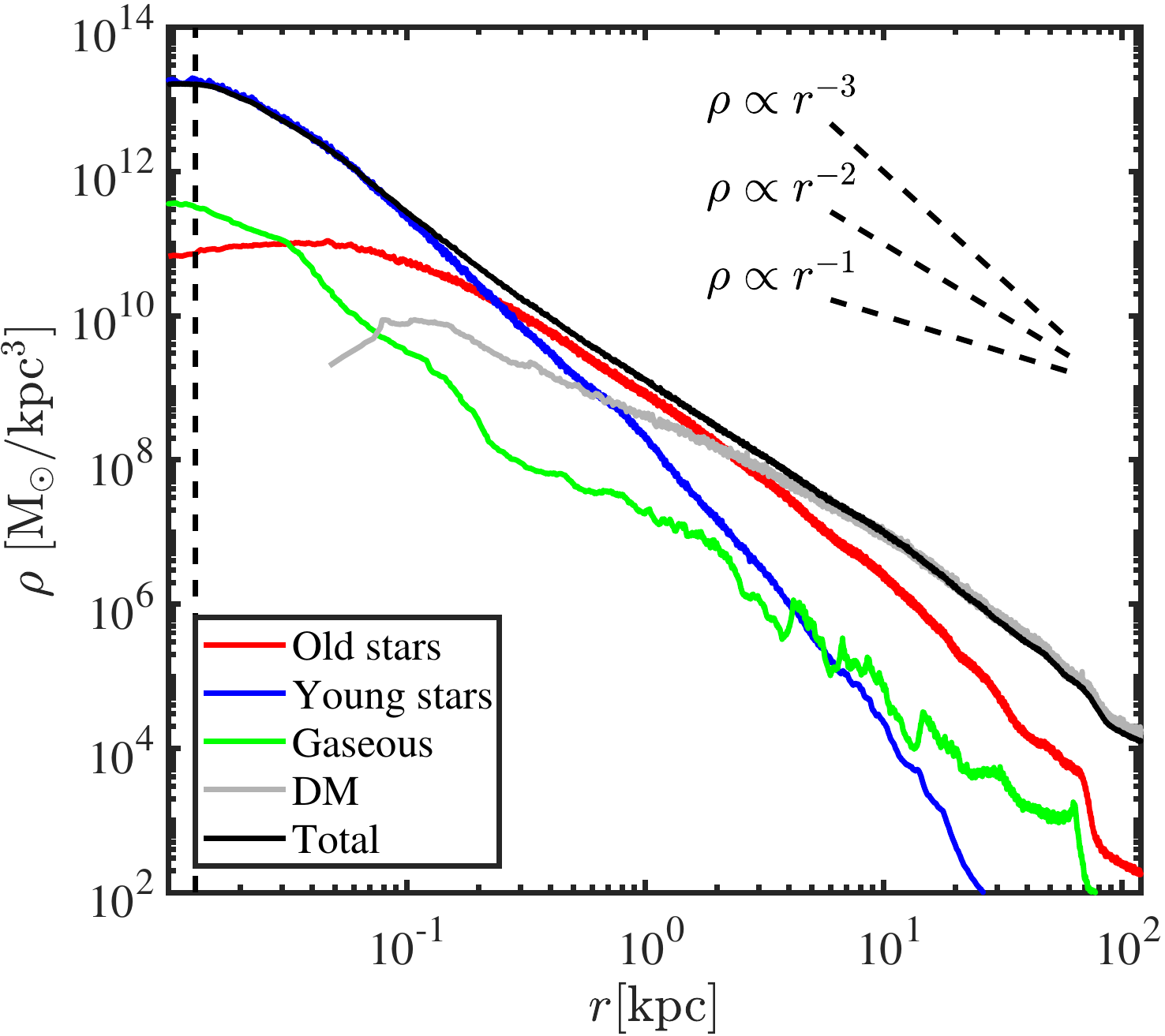}
    \caption{The density profiles of old stars (present before the first pericentric passage),
    young stars, gas, dark matter and all matter combined for the $1$ Gyr old 
    remnant. The profile of dark matter extends only down to $r\sim50$ pc because of low number of particles in the inner region.
    For reference three power-law density profiles with $r^{-1}$, $r^{-2}$ and $r^{-3}$ are also 
    shown. The black dashed line shows the baryonic gravitational
    softening of $13$ pc.}
    \label{fig:density}
\end{figure}

We show in Fig. \ref{fig:density} the radial density profile of the matter components for the remnant $1$ Gyr after the coalescence.
The combined density of all matter is dominated in the central region of the remnant
by luminous matter, where most of the contribution comes from stars.
Young stellar particles, formed after the first pericentric passage at approximately $t\sim 0.7$ Gyr, 
contribute the majority of the stellar density in the inner $220$ pc. These stars have either formed out of the disturbed gaseous disc
during the interaction of the galaxies, 
or formed during the funneling of gas into the central region. Older stars 
dominate the stellar density for most of the radial extent of the remnant galaxy.

The gaseous density
is approximately two orders of magnitude lower than the total mass density for the entire extent of the galaxy
with a profile similar to the stellar density profile.
The dark matter density has a shallower profile compared to the luminous components, and dark matter
begins to dominate the total density from $2.5$ kpc outwards. The dark matter profile steepens beyond $12$ kpc,
which reflects the distribution of dark matter in the initial conditions where the scale lengths of the Hernquist 
halo profiles were $15$ kpc. 

For reference, Fig. \ref{fig:density} shows three power-law
slopes $\rho \propto r^{-\gamma}$ with power-law indices $1, 2$ and $3$, 
where the $\gamma=2$ corresponds to an isothermal profile. The radial profile of the total density in the merger remnant is best
fit with a power-law index $\gamma=2.35$.
The radial lower limit of the fit was set at $2.8$ times the gravitational softening at $r=36.4$ pc, corresponding the limit of the
force being exactly Newtonian, and the upper limit to the virial radius $r_\mathrm{vir}\approx 188$ kpc, where the mean density reaches $200$ times
the critical density.
Earlier studies of binary mergers and the cosmological evolution of elliptical galaxies have shown 
similar results for the total density profile. For example 
\citet{2013ApJ...766...71R, 2017MNRAS.464.3742R} report a distribution of $\gamma$ values peaking at $\gamma\approx2.0$--$2.1$
tending towards slightly lower values for lower mass galaxies. \citet{2017MNRAS.464.3742R} also note 
that ignoring supermassive black holes and the associated feedback in the simulation models leads to steeper values of $\gamma$ 
for the total density.
Observations of early-type galaxies such as the SLACS gravitational lensing 
study \citep{2006ApJ...638..703B, 2008ApJ...682..964B},
report also nearly isothermal profiles with $\gamma=2.01$--$2.085$ for the total density of early-type galaxies 
\citep{2006ApJ...649..599K, 2009ApJ...703L..51K} out to tens of effective radii \citep{2007ApJ...667..176G}.
The density slopes of both observed and simulated individual galaxies have typically a $ 1\sigma$ intrinsic scatter
of the order of $\pm0.1$--$0.2$, while the range of $\gamma$-values for individual ETGs varies as much as $\pm0.5$.
The fitted power-law index is therefore in good agreement with the upper envelope of observed $\gamma$-values. We thus conclude
that the simulated Antennae merger results in a fairly representative ETG and we would most likely 
obtain a slightly shallower density power-law index by including also central supermassive black holes in our simulations.

\subsection{Oxygen abundance}

\begin{figure*}
\includegraphics[width=\textwidth]{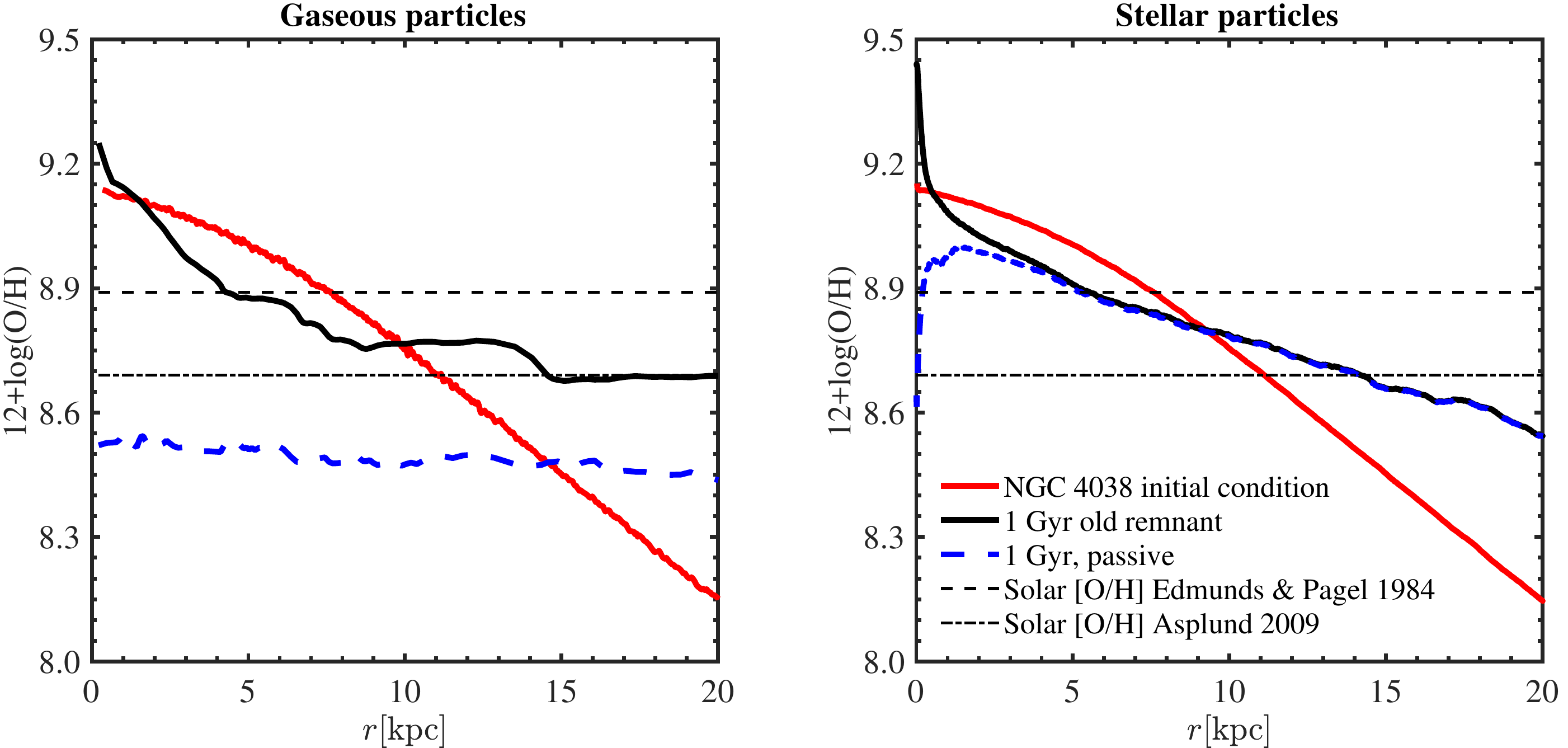}
    \caption{Gaseous (left) and stellar (right) oxygen abundance of the $1$ Gyr old remnant (black) compared to the
    passively evolved metallicity obtained from the initial conditions (dashed blue, see text for details) and
    the initial abundance of the NGC 4038 progenitor (red). The observed solar 
    abundances are also shown for reference.}
    \label{fig:remnant_metals}
\end{figure*}

The intrinsic radial oxygen abundances in the $1$ Gyr old remnant are shown in Fig. \ref{fig:remnant_metals} for the gaseous particles and 
the stellar particles separately, and are compared with the initial abundance of the NGC 4038 progenitor. 
Again we only consider the intrinsic metallicity values of the particles
as given directly by the simulation.
To separate the contribution from diffusion and stellar enrichment versus 
passive mixing due to dynamical evolution, we also show the $[\mathrm{O}/\mathrm{H}]$ profile which would result from a purely
passive redistribution of the particles. This is done by tracing the initial abundances
of the particles in the remnant back to the initial conditions
and calculating the radial abundance as if it would have evolved passively.

For gas, both the enriched and the passively evolving $[\mathrm{O}/\mathrm{H}]$ profiles in Fig. \ref{fig:remnant_metals} show that 
the initial metallicity gradients of the progenitor discs 
get mostly erased when the particles in the gaseous discs get redistributed during the 
merger. 

Merging systems are typically outliers in observed metallicity relations.
The metallicity profiles of interacting galaxies have been previously studied extensively using both simulations (e.g. 
\citealt{2006A&A...459..361P, 2010ApJ...710L.156R, 2010A&A...518A..56M}) and also observationally 
(e.g. \citealt{2008ApJ...674..172R, 2008AJ....135.1877E, 2010ApJ...721L..48K, 2014A&A...563A..49S}), where the 
increasing capabilities of extensive integral field spectroscopic surveys have been particularly important. 
The general consensus is that the metallicity gradient flattens, as the metal-poor gas in the outer 
regions of the progenitors gets mixed up with the more metal-rich gas in the inner regions of the interacting galaxies. 
The majority of this dilution is due to purely dynamical effects as already noted in \citet{2010ApJ...710L.156R}, who did not 
include active star formation in their study. In agreement with previous studies, the simulated Antennae remnant 
in both the passively traced and the full enrichment model has an $[\mathrm{O}/\mathrm{H}]$ value below the initial metallicity 
of the progenitor galaxy for most of observationally accessible inner region of the galaxy.

As already pointed out in studies involving larger simulation samples, the enrichment through star formation and 
stellar feedback has a significant role in determining the spatial distribution of metals in both the gaseous and 
stellar components (e.g. \citealt{2010A&A...518A..56M, 2011MNRAS.417..580P, 2012ApJ...746..108T}).
Our merger simulation employing active metal enrichment results in an increase of $[\mathrm{O}/\mathrm{H}]$ by $0.2$ dex
over the entire radial range compared to the passively evolving simulation. We note that similar results have also been reported in \citet{2012ApJ...746..108T},
where they found a mean nuclear $[\mathrm{O}/\mathrm{H}]$ enhancement due to enrichment already during 
the pre-coalescence phase by at least $0.2$ dex.
The passive mixing of the gaseous gradients dominates the redistribution in the outer parts, where
the abundance profile remains flat.
However, when active enrichment is considered the metallicity
in the innermost region of the remnant is drastically different from the purely
dynamically evolving $[\mathrm{O}/\mathrm{H}]$. Firstly, the radius outside of which the 
metallicity switches from underabundant to enhanced, with respect to the initial conditions, shifts $5$ kpc
inwards due to the overall $0.2$ dex radial enrichment. Secondly,
within the innermost $8$ kpc the oxygen abundance shows a steepening
gradient towards the centre of the remnant. The abundance actually
exceeds the initial abundance within the innermost $1.5$ kpc.
This increase in metallicity is due to the central
starburst resulting in enhanced star formation and enrichment. 

According to the passively distributed stellar metallicity in the right hand panel of Fig. \ref{fig:remnant_metals}, 
the inner $1.5$ kpc would have an inwards decreasing abundance,
whereas the result from the full feedback implementation results in a super-solar enhancement of over $0.5$ dex.
The difference can be explained with the properties of the central star-forming gas in the two models.
The stellar mass within the inner region is dominated by young stars out to $220$ pc (see Fig. \ref{fig:density}),
where the metallicities of the young stellar particles are defined by the star-forming gas.
The central depletion in the passive stellar metallicity follows from the flattened profile of
the passively evolved gaseous metallicity, whereas the enriched model shows an enhanced metallicity resulting from the enriched gas 
(see left hand side of Fig. \ref{fig:remnant_metals}).
However, the overall stellar metallicity of the remnant is for the majority of the radial range set by dynamical evolution,
as is shown by the overlap of the passively evolved and enriched profiles.
The $[\mathrm{O}/\mathrm{H}]$ gradient flattens from the initial $0.06$ dex$/$kpc to $\lesssim 0.028$ dex$/$kpc (in the linear region $r\gtrsim 1$ kpc) 
still maintaining a clearly decreasing trend towards outer radii. This is in agreement with observations of metallicities in early-type 
galaxies with similar stellar masses as the Antennae remnant (e.g. \citealp{2010MNRAS.407..144T,2012MNRAS.426.2300L}). 
In general observed lower mass late-type galaxies exhibit steeper abundance gradients, whereas for ETGs with $M_*\gtrsim2\times10^{10}$ M$_{\sun}$ 
the gradients flatten towards larger masses \citep{2010MNRAS.407..144T}, in good agreement with our simulation results.

\subsection{Velocity dispersion}

\begin{figure}
\centering
\includegraphics[width=\columnwidth]{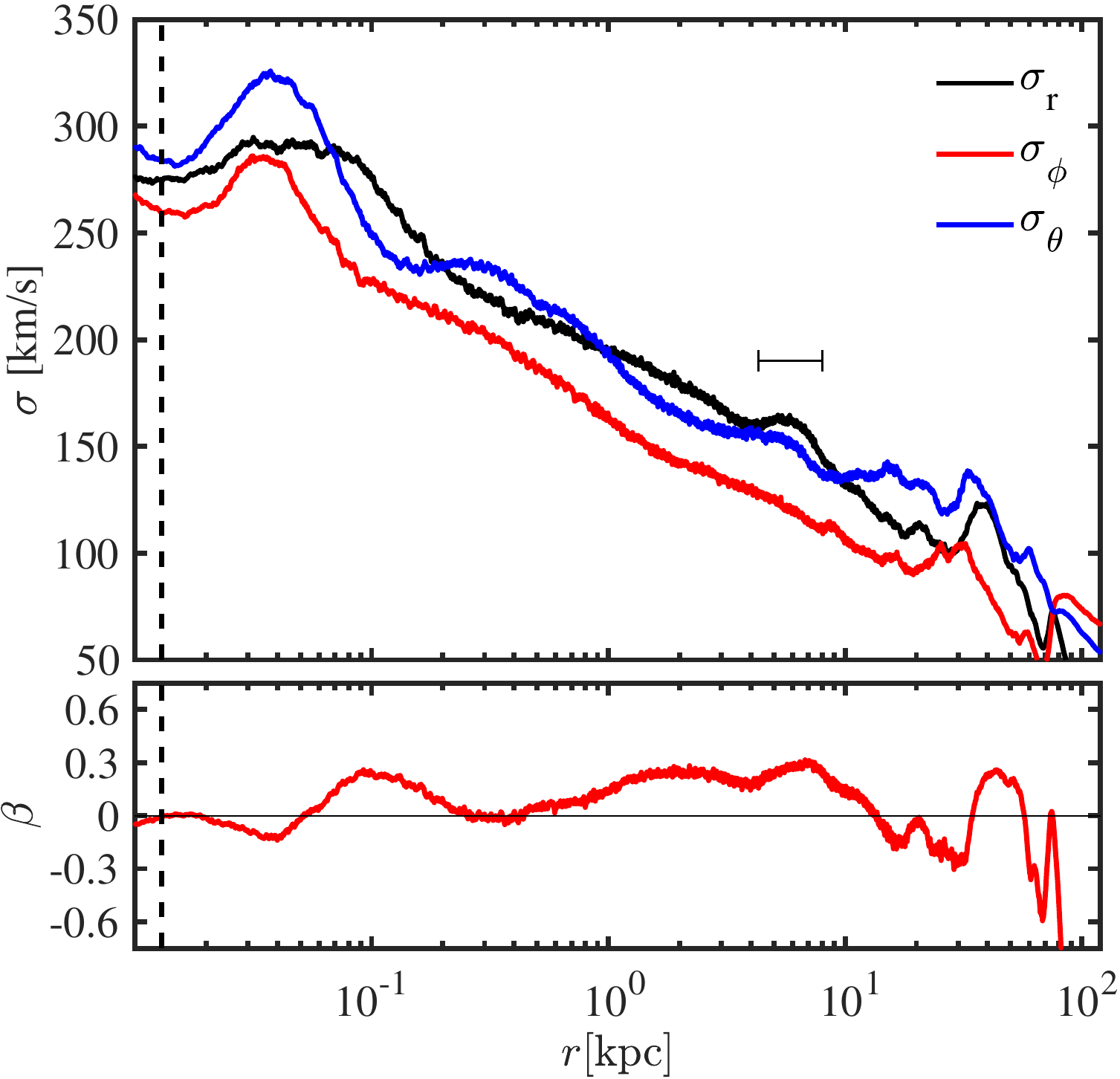}
    \caption{Top: the stellar velocity dispersion profile separated into the intrinsic radial (black) and the two tangential (red and blue)
    components. The bar indicates the position of the $\sigma$-bump (see text for details).
    Bottom: the anisotropy profile calculated from the radial and
    tangential dispersion profiles as in Eq. \ref{eq:anisotropy}.
    The dashed line in both of the panels indicates the baryonic gravitational softening length of 
    $13$ pc.}
    \label{fig:sigma_profiles}
\end{figure}

An important observable that can be derived from most surveys is the velocity dispersion.
Planetary nebulae and globular clusters can be observed at radii exceeding the effective radius, where gas is seldom present in substantial quantities.
Here we analyse the intrinsic dispersion profiles, and revisit in Sec. \ref{section:kinemetry} the subject by studying 
the projected line-of-sight velocity dispersion. 
In Fig. \ref{fig:sigma_profiles} we present the intrinsic radial and tangential 
velocity dispersions, $\sigma_r (r)$, $\sigma_\phi (r)$ and $\sigma_\theta (r)$
for the stellar particles  as a function of radius.

The Antennae remnant is fairly isotropic, as the different components
of the velocity dispersion remain within $50$ km$/$s of each other for the entire radial range.
The dispersion profiles reach their maxima of $\sigma_{r,\mathrm{max}}=294$ km$/$s, 
$\sigma_{\phi,\mathrm{max}}=285$ km$/$s and $\sigma_{\theta,\mathrm{max}}=326$ km$/$s
at $\sim35$ pc and decrease fairly smoothly out to $r\sim30$ kpc. 
The profile is in agreement with the line-of-sight velocity dispersion of local ETGs, where the range of values is typically from $50$
to $300$ km$/$s measured within the effective radius \citep{2013MNRAS.432.1709C}.
From $r\approx40$ kpc outward there are
multiple stellar streams, seen in projection as typical merger-induced shells, 
which are seen as locally oscillating values in the dispersion profiles.

The relation between the radial and tangential dispersion components can be more quantitatively expressed with 
the radial anisotropy parameter defined as
\begin{equation}\label{eq:anisotropy}
 \beta (r) \equiv 1-\frac{\sigma_{\phi}^2(r)+\sigma_{\theta}^2(r)}{2\sigma_{r}^2(r)},
\end{equation}
which is shown in the bottom panel of Fig. \ref{fig:sigma_profiles}.
The anisotropy parameter may range from $\beta \to -\infty$ for a purely rotating 
stellar system to $\beta=1$ for purely radial dispersion. 
The anisotropy parameter of the stellar component stays between $-0.3<\beta<0.31$ for most of the
radial range, tending towards slightly radially biased values at intermediate radii of $0.05$--$13$ kpc.
This is in agreement with other simulation studies that found that dissipational mergers tend to produce 
fairly isotropic, radially biased merger remnants \citep{2006ApJ...650..791C}.

In the region of $4$--$8$ kpc a slight $10$--$20$ km$/$s increase in the dispersions of the $\sigma_r (r)$ and $\sigma_\theta (r)$ profiles
can be seen, indicated in Fig. \ref{fig:sigma_profiles} with a horizontal bar. This type of a feature has been identified in the literature
as a '$\sigma$-bump', which differs fundamentally from the oscillating nature of the signature
caused by the outer shell structures \citep{2014ApJ...783L..32S}. The shells can be
seen as enhancements in the radial density profile, whereas the $\sigma$-bumps have not been associated
with any clear structures in galaxies as is also the case in the Antennae remnant (see Fig. \ref{fig:density}).
Strong $\sigma$-bumps have also been detected only in the stellar component of simulated merger remnants.

\section{Observed photometry of the merger remnant}\label{section:photometric_properties}

\subsection{Light profile}

\begin{figure}
    \includegraphics[width=\columnwidth]{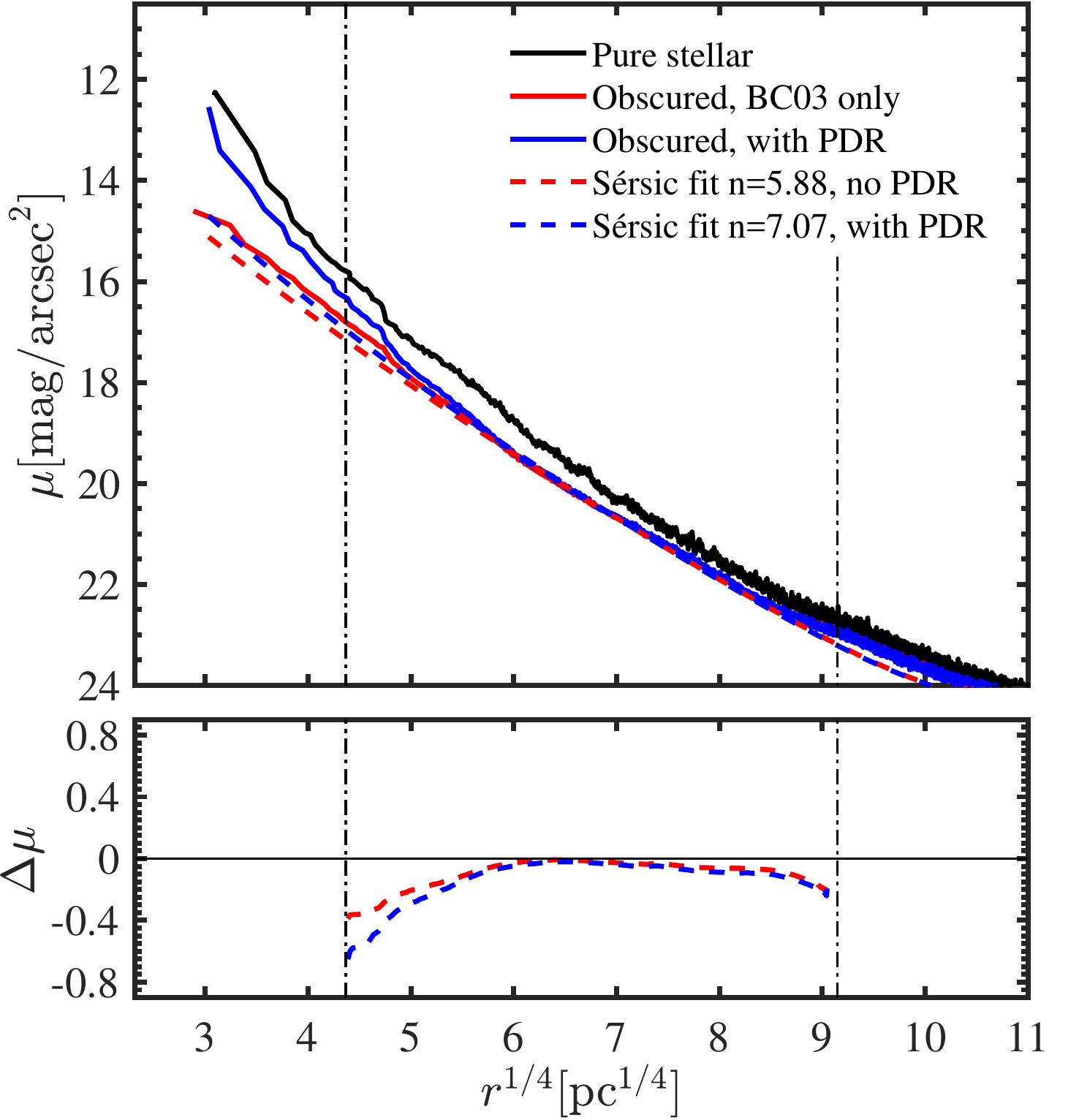}
    \caption{The unobscured (black solid line) and obscured projected surface 
    brightness profiles in the $r$-band mock-image of the $1$ Gyr old merger remnant
    with resampling of young stars (blue solid line) and without (red solid line), overlaid with the
    respective best fit S\'{e}rsic profiles with dashed lines.
    The bottom panel shows the $data-fit$ residuals for the profile with 
    resampling (blue) and without (red) in the fitted region. 
    The vertical dot-dashed lines show
    the inner and outer fit limits given by the $2.5$ arcsec mask and the median SDSS $r$-band sky value at 
    $\sim23$ mag$/$arcsec$^2$ (Pawlik et al., in prep).}
    \label{fig:light_profile_with_fits}
\end{figure}

The unobscured (purely 
stellar) and obscured (including dust in the ISM and the PDRs) 
projected surface brightness profiles of the $1$ Gyr old merger 
remnant are shown in Fig. \ref{fig:light_profile_with_fits}. The profiles are calculated from 
the SDSS equivalent $r$-band image of the remnant as it would be seen in the sky, produced with \skirt\ assuming a distance of $30$ Mpc. 
For reference we show the dusty profiles of both the image with the resampling of the young stellar component (PDR) and 
without the resampling (BC03 only). The unobscured remnant shows a cuspy 
profile towards the centre, caused by the concentrated star formation in the central region.
The inclusion of dust obscuration suppresses the surface brightness within the entire radial range
of $r<6.5$ kpc, which is best seen in the dusty BC03 profile which excludes emission from the very youngest stellar component.
The strongest obscuration of up to $3$ mag$/$arcsec$^2$ is seen within the central $2$ kpc of the dusty BC03 profile, which 
contains $\sim 3.6\times 10^8$ M$_{\sun}$ of gas.
We note that observed ETGs evolving towards the red sequence have been found to include
significant amounts of dust and ongoing star formation (e.g. \citealt{2012MNRAS.419.2545R}).
 Even though the remnant is clearly an elliptical-like galaxy, at an age of $1$ Gyr it is still transitioning towards a classical red and dead ETG.

The resampling procedure introduces $1.7 \times 10^7$ M$_{\sun}$ of star-forming regions (PDR-particles)
in the inner galaxy. Including the SED of the young stellar component increases the surface brightness  
up to the level of the unobscured profile in the central region of the galaxy. 
The properly processed young stellar disc is therefore more alike the unobscured BC03 image within a cuspy central region,
whereas the remainder of the light profile follows the dusty BC03 with reduced brightness wherever dust is present.

\subsection{Fitting the photometric profile}

Next we use \galfit\ to find the radial surface brightness profiles
of the $1$ Gyr old remnant presented in Fig. \ref{fig:light_profile_with_fits}. We fit the
profiles to the $r$-band image following \citet{2013MNRAS.432.1768K}, 
who fit either a single component or a two component (disc $+$ bulge) S\'{e}rsic 
profile where the innermost $2.5$ arcsec is masked. The radial surface 
brightness profile of an ETG can be expressed 
using the S\'{e}rsic profile
\begin{equation} \label{eq:brightness}
 I(r)=I_{\rm{e}} \exp{ \left\{ b_n\left[\left(\frac{r}{r_{\rm{e}}}\right)^{1/n}-1\right]\right\}},
\end{equation}
where $I_{\rm{e}}$ is the surface brightness at the effective radius $r_{\rm{e}}$, $n$ is the 
S\'{e}rsic index and $b_n$ is a shape-factor dependent on $n$ as 
$b_n\approx 1.9992n-0.3271$ for $0.5<n<10$ \citep{1989woga.conf..208C}. The 
projected surface brightness profile in mag$/$arcsec$^2$ 
can be derived from Eq. \ref{eq:brightness} in the Vega-system as
\begin{equation} \label{eq:projected_brightness}
 \mu(r)=\mu_{\rm{e}}+\frac{b_n}{\ln{10}}\left[\left(\frac{r}{r_{\rm{e}}}\right)^{1/n}-1\right],
\end{equation}
where $\mu_{\rm{e}}=-2.5\log{I_{\rm{e}}}$ is the surface brightness at the effective radius. \galfit\
provides the best-fit values of $r_{\rm{e}}$, $\mu_{\rm{e}}$ and $n$ for the S\'{e}rsic-profile assuming the Vega-system.
For comparison to the \atlas\ data which uses asinh-magnitudes, the projected surface 
brightness profiles are expressed in Fig. \ref{fig:light_profile_with_fits} in the asinh-system as
\begin{equation} \label{eq:projected_brightness_asinh}
 \mu(r)=-\frac{2.5}{\ln{10}}\left[\mathrm{asinh}\left(\frac{f/f_{\rm{0}}}{2b}\right)+\ln{b}\right]
\end{equation}
where $f$ is the flux, obtained for the S\'{e}rsic-profile from Eq. \ref{eq:brightness}, $f_{\rm{0}}$ is the zero point flux $3631$ Jy and $b$ 
is a filter-dependent constant\footnote{http://classic.sdss.org/dr7/algorithms/fluxcal.html}. 
We fit both the image with and without the resampling of the
young stellar component in the asinh-system.

Observations have shown that ETGs typically exhibit S\'{e}rsic 
indices close to the de Vaucouleurs value $n\sim4$ \citep{2009ApJS..182..216K}. If a cusp is apparent in 
the surface brightness profile of an ETG, an additional disc-like light
profile  can be fitted with a S\'{e}rsic index $n\sim1$
(see e.g. \citealt{2009ApJS..181..135H}). The central cusp is a characteristic property 
stemming from the central merger induced starburst. 
However, the innermost regions of ETGs are usually masked out which may exclude the inner disc 
from the fitted region.
Here, following \atlas\ we use an equivalent mask of $2.5$ arcsec in the inner galaxy
and limit the fit to a median SDSS $r$-band sky value of $\sim23$ mag$/$arcsec$^2$
at approximately $6.6$ kpc. 

We show the best fit surface brightness profiles
in Fig. \ref{fig:light_profile_with_fits} with the residuals for both of the profiles with and without resampling of the young stars.
The fits are single component S\'{e}rsic profiles, with the fitting performed between the two vertical dot dashed lines, thus excluding 
the very central parts and the outer low surface brightness region. 
We find $r_{\rm{e}}=11.07$ arcsec $=1.61$ kpc, $\mu_{\rm{e}}=19.81$ mag$/$arsec$^2$ and $n=7.07$ for the profile with the resampling of young stars
and $r_{\rm{e}}=13.51$ arcsec $=1.97$ kpc, $\mu_{\rm{e}}=20.26$ mag$/$arsec$^2$ and $n=5.88$ for the profile without resampling. 
The derived S\'{e}rsic indices are within the range of typical values for local ETGs with
dynamical masses $> 4\times 10^{10}$ M$_{\sun}$ where the distribution peaks at $n\approx3-3.5$ \citep{2009ApJS..182..216K, 2013MNRAS.432.1768K}. 
However, not using the inner mask might enable the fitting of two component profiles, thus decreasing the outer S\'{e}rsic index and 
making the result more similar to the earlier simulated results of \citet{2008ApJ...679..156H,2009ApJS..181..135H,2009ApJS..181..486H}.
The effective radii are towards the lower end of the observed distribution (peak at $r_{\rm{e}}\approx3$--$4$ kpc), 
but still within the observed range \citep{2013MNRAS.432.1768K}. Within the fitting region both the profiles with and without
resampling of young stars are quite similar, resulting in similar S\'{e}rsic fits. In general the fits agree with the 
surface brightness profiles for most of the fitted region (i.e. $\sim 1$--$6.5$ kpc) and the major disagreements between the fits and
the surface brightness profiles appear in the inner regions, with an increasing contribution from the central star-forming disc.

\galfit\ also provides the axis ratio $b/a$ of the fitted ellipsoids at the effective radius. With a stellar mass of 
$\sim 10^{11}$ M$_{\sun}$, our remnant is in the intermediate mass range of ETGs
and should therefore not show a high ellipticity in contrast to fast
rotating lower mass elliptical galaxy. As indicated by the intrinsic shape of the 
stellar component in Sec. \ref{section:intrinsic_structure} (see also Sec. \ref{section:atlas_counterpart}), 
the remnant shows only a slight elliptical shape, with $\epsilon=1-b/a=0.1-0.2$.
The single component fit results in ellipticities at the effective radius of $\epsilon_{\rm{e}}=0.122$ and $\epsilon_{\rm{e}}=0.120$, compared to
the intrinsic ellipticity of $\epsilon\approx0.09$ at the same radius (see \ref{section:intrinsic_shape}). 
The shape of the remnant is therefore consistent with being only slightly elliptical both intrinsically
and also when observational and projection effects are taken into account.

\subsection{Colour evolution}\label{section:colour_evolution}

\begin{figure}
    \includegraphics[width=\columnwidth]{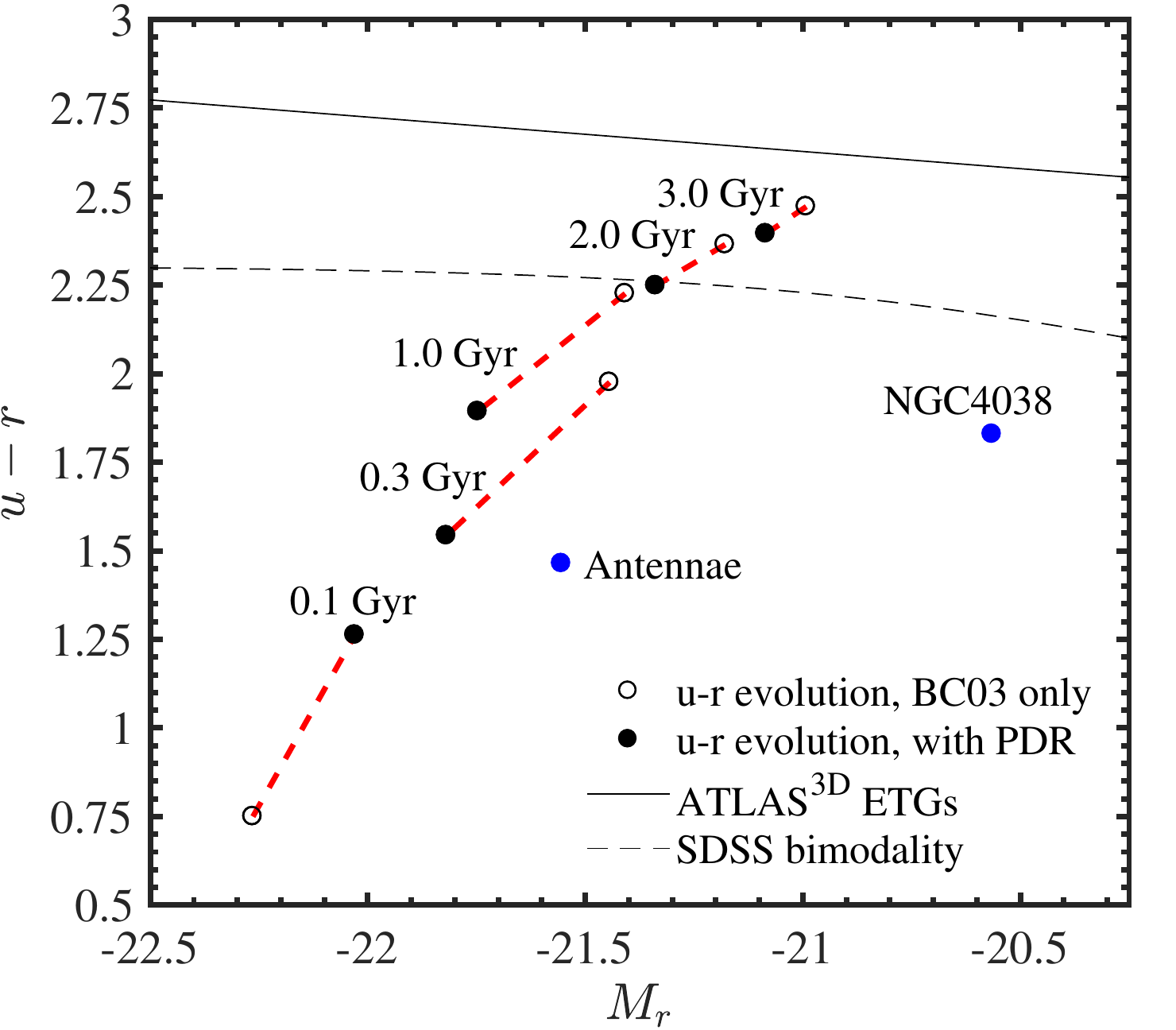}
    \caption{The evolution of the $u-r$ colour of the simulated Antennae merger as a function 
of the absolute $r$-band magnitude. 
The blue dots show the colours of the simulated NGC 4038 progenitor as well
as the best match Antennae including the PDR resampling. The solid black points show the magnitude-colour
evolution for the model including the resampling of the young stellar component and the open circles 
show the corresponding evolution for the BC03 only model without young star resampling. 
The colours are calculated for the remnant approximately $0.1$, $0.3$, $1$, $2$, and $3$ Gyr 
after the final coalescence and the corresponding two data points are always connected by a red dashed line.
The black solid line shows the best-fit colour-magnitude relation for the local ETGs in 
the \atlas\ sample \citep{2011MNRAS.413..813C}
and the dashed black line shows the division between blue and red galaxies in the SDSS survey \citep{2004ApJ...600..681B}.}
    \label{fig:colours}
\end{figure}

The evolution of the $u-r$ colour of the remnant is shown in Fig. 
\ref{fig:colours} as a function of the absolute $r$-band magnitude. The colour
evolution is studied until $3$ Gyr after the coalescence.
The inclusion of the resampling of the young stars provides extra blue light and shifts the remnant
colours bluewards, while the $r$-band magnitude decreases mildly due to
additional re-emission from dust in the volume around the star-forming regions.
As a reference we show the colour of the simulated progenitor NGC 4038 (after $500$ Myr of isolated evolution) 
and the best match Antennae in Fig. \ref{fig:colours}, both employing the resampling of the young stellar component.

The Antennae appears $\sim0.37$ mag bluer in the $u-r$ colour and $\sim1$ mag brighter in the $r$-band compared to the
progenitor galaxy. After the second passage $\sim 40$ Myr before the best match, the 
star formation rate has increased from $\le 10$ M$_{\sun}/$yr 
to values above $20$ M$_{\sun}/$yr.
The merger appears its bluest and brightest in the $r$-band
right after the final coalescence of the nuclei. 
As the remnant matures after coalescence it gets dimmer and redder as the starburst 
subsides and the stellar population evolves predominantly passively. 

Immediately after the coalescence, the residual star formation 
and additional dust in the starburst region can still be seen as a reddening
of the colour of the $100$ Myr old remnant when comparing the \mappings-included and BC3 only colours. 
After the $100$ Myr data point, the remainder of the
colour evolution follows a qualitatively similar path towards the red sequence regardless of the chosen stellar library.
However, for intermediate ages of $\sim 0.3-1.0$ Gyr the additional blue light in the \mappings\ model 
makes the evolution of the remnant up to $0.5$ mag bluer in the  
$u-r$ colour when compared to the BC03 only colour model.
The asymptotically decreasing star formation rate reaches its minimum of $\sim 0.8$ M$_{\sun}/$yr after at a remnant age 
of $2.5$ Gyr. At this age
the contribution from young stars and therefore the additional \mappings\ flux becomes negligible.
After this the spectral evolution is again governed by the increasing age of the stellar population
and dominated by the BC03 SED, which drives the colour evolution of the old remnant galaxy. 

As a comparison with observations we show the best-fit $u-r$ vs. $M_r$ colour --
magnitude relation (CMR) for ETGs in the \atlas\ sample with a
slope $\eta=-0.097$ and zero-point $(u-r)_{\rm{0}}=2.53$ in asinh-magnitudes \citep{2011MNRAS.413..813C}.
In order to express the bimodality between the mainly blue spirals and red ETGs in the colour--magnitude plane, we also show the
optimal division line between the two regimes for local SDSS galaxies with $-23.5 \le M_r \le -15.5$ \citep{2004ApJ...600..681B}.
Immediately after the coalescence, the merger remnant appears considerably bluer when compared to local early-type galaxies.
The $u-r$ colour increases as the remnant dims more rapidly in the $u$-band, and the $u-r$ colour
approaches the observed $u-r \gtrsim2.5$ mag value
typical for ETGs with $-22 \lesssim M_r\lesssim -18$. At an
age of $\gtrsim2.5$ Gyr the remnant crosses from the blue cloud to the red sequence, 
yet it still resides below the local best-fit CMR.
\citet{2011MNRAS.413..813C} note that the observed colours of massive ETGs in the \atlas\ sample
are close to the best-fit CMR, while some of the lower mass ETGs are still only transitioning towards the relation.

\section{Observed kinematic structure of the merger remnant}\label{section:kinemetry}

\subsection{2D line-of-sight kinematics}

Next we compute the 2D stellar kinematics of the $1$ Gyr old remnant in the inner galaxy 
covering approximately $1$--$2$ effective radii. 
In calculating the line-of-sight velocity distribution (LOSVD) we closely follow the
procedure presented in \citet{2014MNRAS.444.3357N}, which was modelled on the method used to
extract LOSVD properties from the observational \atlas\ sample.  
To reduce effects from the particle mass resolution, the method replaces each 
stellar particle with $10$ pseudo-particles with a two-dimensional Gaussian 
distribution and equivalent line-of-sight velocity. The pseudo-particles are distributed with a 
standard deviation of $13$ pc, equivalent to the gravitational softening length, which emulates also the seeing effect. 
 These particles are then 
binned on a regular grid of pixels. The pixels are further grouped using the Voronoi tesselation method presented in 
\citet{2003MNRAS.342..345C} to produce an 
irregular grid with roughly 
equal amount of particles in each bin. Using the maximum likelihood method we then calculate the mean line-of-sight velocity (LOSV) 
$\left\langle V \right\rangle$, velocity dispersion $\sigma$, and the coefficients $h_3$ and $h_4$ representing the third and fourth 
moments of Gauss-Hermite series fit to the LOSVD. 

\begin{figure*}
\centering
\includegraphics[width=0.9\textwidth]{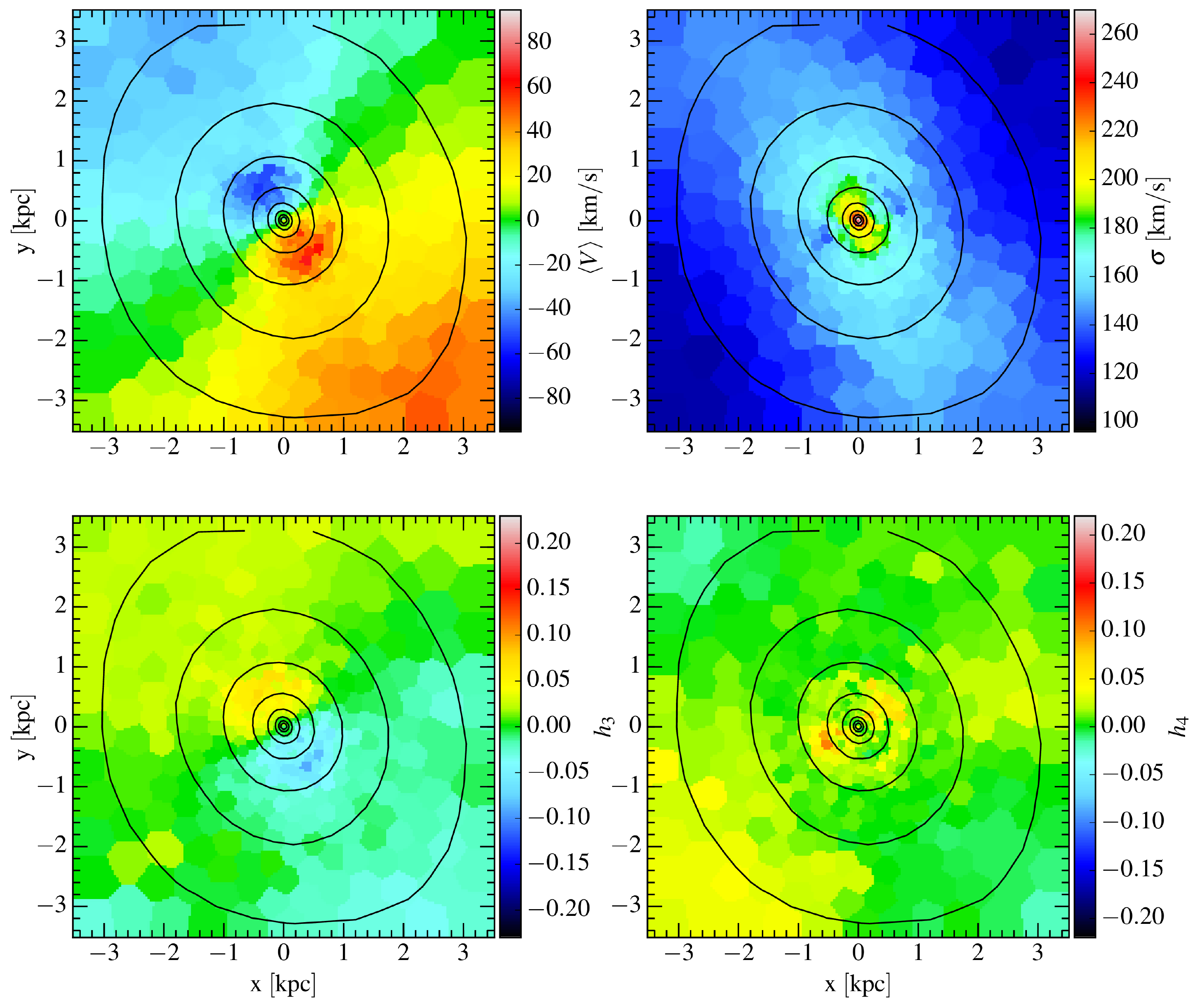}
    \caption{Mean line-of-sight velocity $\langle V \rangle$ (top left), velocity dispersion $\sigma$ (top right), 
    and the higher order Gauss-Hermite coefficients $h_3$ (bottom left) and $h_4$ (bottom right) in the inner $3.5$ kpc 
    (out to $\sim 2$ effective radii) of the
    Antennae remnant as it would be seen on the sky $1$ Gyr in the future. The contours show the 
    distribution of flux (mass) in $1$ mag intervals. The range of the colour bars is adjusted to be comparable with the 
    analysis presented in Sec. \ref{section:atlas_counterpart}.}
    \label{fig:velocity_distribution}
\end{figure*}

Figure \ref{fig:velocity_distribution} shows all the fitted stellar LOSVD 
properties of the $1$ Gyr old remnant within the innermost $3.5$ kpc which encompasses 
approximately the area within two effective radii. The LOSVD 
has a spread of approximately $\left\langle V \right\rangle=\pm 60$ km$/$s, 
with a maximum velocity dispersion of $\sigma=270$ km$/$s found in the centre.
The rotational velocity in the central region increases by $\sim15$ km$/$s 
as the remnant evolves to an age of $3$ Gyr. This is caused by  
the contribution from newly formed stars in the central rotating disc, 
which are added during the late evolution of the merger remnant. 

Next we compare the central region of the remnant with 
observed ETGs. The well known fundamental plane (FP) of ETGs shows a relatively tight relation with small scatter
between the velocity dispersion within the effective radius $\langle\sigma_{\rm{e}}\rangle$, the effective radius $r_{\rm{e}}$ and 
the surface brightness within the effective radius $\langle \mu_{\rm{e}}\rangle$ 
\citep{1976ApJ...204..668F,1987ApJ...313...59D,2003AJ....125.1866B, 2006MNRAS.366.1126C}.
For our merger remnant the mean velocity
dispersion within the effective radius $r_{\rm{e}}\approx 1.6$ kpc is $\langle\sigma_{\rm{e}}\rangle\approx180$ km$/$s and the mean surface
brightness in the $r$-band within the effective radius is 
$\langle \mu_{\rm{e}}\rangle\approx18.1$ mag$/$arcsec$^2$. Comparing these values to the FP fit at a corresponding $r$-band, given
in \citet{2003AJ....125.1866B} for nearly $9000$ ETGs in the SDSS, we find that our $1$ Gyr old merger
remnant coincides almost exactly with the observed $r_{\rm{e}}$--$\langle\sigma_{\rm{e}}\rangle$--$\langle \mu_{\rm{e}}\rangle$ relation. 
However, while only $0.02$ dex above the best fit FP, the value for the Antennae remnant is located in the small--bright--low dispersion
end of the individual observations, due the bright star-forming central disc. 

\begin{figure*}
\centering
\includegraphics[width=\textwidth]{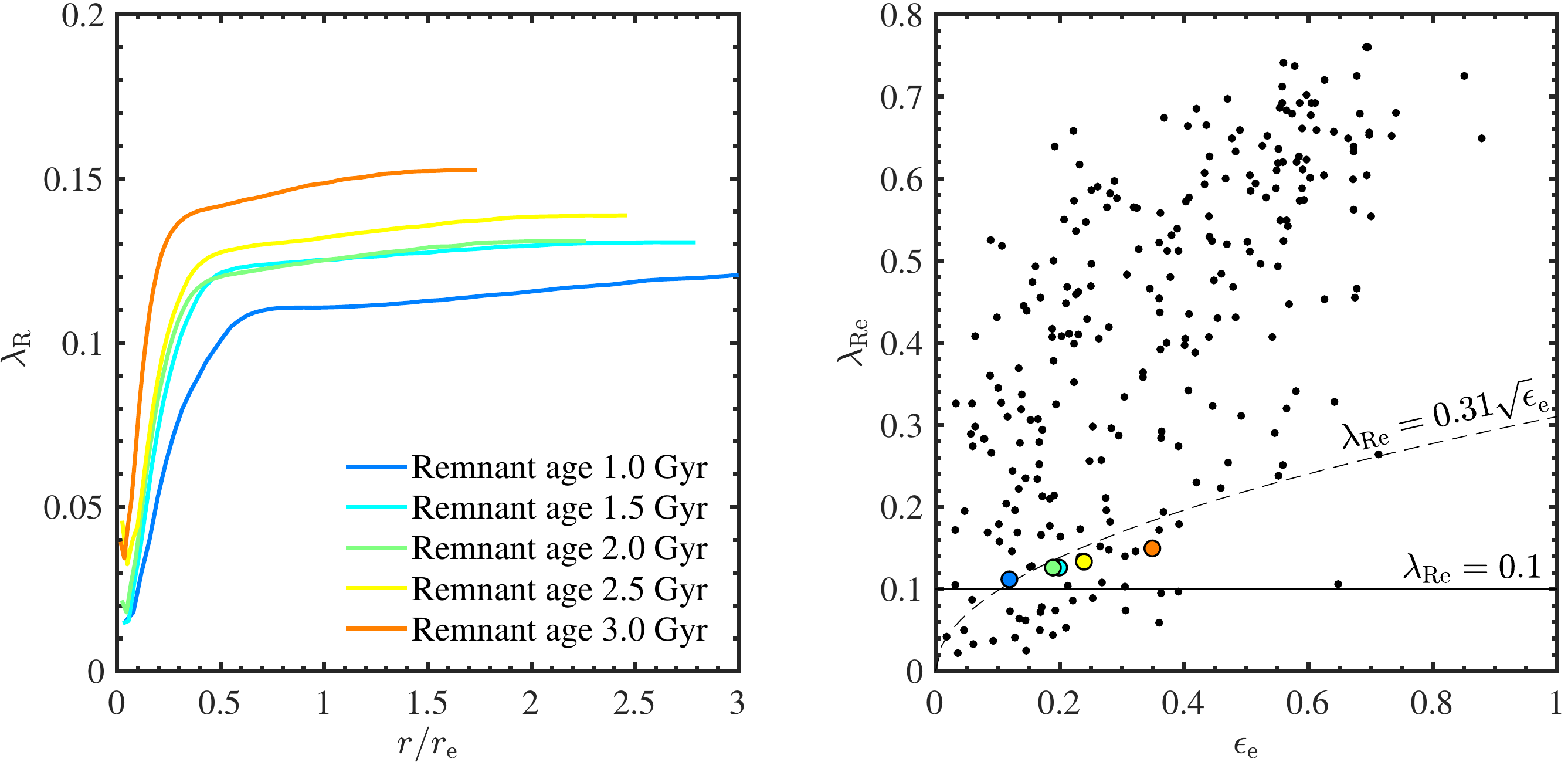}
    \caption{Left: The $\lambda_{\rm{R}}$-parameter as a function of radius scaled by
    the $r$-band effective radius. The different colours indicate the age of 
    the remnant from $1$ Gyr to $3$ Gyr and the curves span approximately $6$ kpc. Right: The values of $\lambda_{\rm{Re}}$
    during the evolution of the remnant as a function of the corresponding ellipticity, 
    compared to the threshold values $\lambda_{\rm{Re}}=0.1$ (dashed) and 
    $\lambda_{\rm{Re}}=0.31\sqrt{\epsilon_{\rm{e}}}$ (solid). Galaxies are classified as slow rotators if they end
    up below the chosen threshold curve. The marker colours are the same
    as in the left hand figure and the black dots show ETGs in the \atlas\ sample \citep{2011MNRAS.414.2923K}.
    Note the different scales on the y-axes.}
    \label{fig:lambda_R}
\end{figure*}

The distribution of $\left\langle V \right\rangle$ in Fig. \ref{fig:velocity_distribution} shows the Antennae remnant as a slightly rotating galaxy with a 
double maximum structure in the distribution of the mean velocity. As identified for many of the ETGs in the \atlas\ sample
\citep{2011MNRAS.414.2923K}, the mean velocity reaches its maximum within the central region, then declines slightly,
and reaches another maximum at a larger radius. The maxima are usually aligned, and the double maximum (2M) feature 
works as one of the indicators for a rotating disc in the centre of the galaxy.
The central $2$ kpc with the rotating stellar
disc exhibits a maximum rotation velocity of $-53/+61$ km$/$s, and the other maximum of $\pm 50$ km$/$s
is situated at $4$ kpc. The $1$ Gyr old remnant would be classified to be a member of the most 
abundant class of ETGs in the \atlas\ sample where $80\%$
of ETGs are classified as ETGs with ordered rotation. On the other hand, in the \atlas\ ETG sample $14\%$ of the galaxies
exhibit the double maximum (2M) feature \citep{2011MNRAS.414.2923K}, making this feature the second most common case after having 
no features at all.

An observed anticorrelation between $\left\langle V \right\rangle/\sigma$
 and $h_3$ corresponds to ordered rotation 
and a disc-like axisymmetric component. This anticorrelation can be 
seen in the central region of the Antennae remnant in Fig. 
\ref{fig:velocity_distribution}, given that the dispersion profile is fairly rotationally symmetric.
The presence of the anticorrelation is in agreement with observations of local ETGs.
For the \atlas\ ETGs \citep{2011MNRAS.414.2923K} a $\left\langle V \right\rangle/\sigma-h_3$ 
anticorrelation is most often detected in galaxies with ordered rotation and galaxies that either have no special features, 
the 2M feature or a twist in the rotational axis of the galaxy. 
In addition, the $\left\langle V \right\rangle/\sigma-h_3$ anticorrelation is usually 
connected to ellipticals with relatively large angular momenta, such as ETGs formed through dissipational mergers 
\citep{2006MNRAS.372..839N, 2014MNRAS.445.1065R}.
Thus, the anticorrelation of $\left\langle V \right\rangle/\sigma$ and $h_3$ can be connected to the presence of gas and the formation 
of gaseous discs in the formation of ETGs, such as is the case for the Antennae remnant.
We also note that the anticorrelated $\left\langle V \right\rangle/\sigma-h_3$ distribution remains strong for the final $2$ Gyr 
in the evolution of the remnant.

The $h_4$ coefficient illustrates a peaked ($h_4>0$) or a flat topped ($h_4<0$) LOSVD, where the value
of $h_4$ roughly correlates with the anisotropy parameter $\beta$ discussed in Sec. \ref{section:kinemetry}
\citep{1993ApJ...407..525V, 1993MNRAS.265..213G}.
The distribution of $h_4$ encompasses predominantly positive values in Fig. \ref{fig:velocity_distribution}, indicating radial anisotropy
consistent with the intrinsic velocity anisotropy in Fig. \ref{fig:sigma_profiles}.

\subsection{Rotation}

In order to evaluate the rotation of the remnant, we compute the $\lambda_{\rm{R}}$-parameter 
\citep{2007MNRAS.379..401E, 2014MNRAS.444.3357N} as
a function of the projected distance of each Voronoi grid cell $i$ from the centre, defined as
\begin{equation}
 \lambda_{\rm{R}}=\frac{\sum_{i=1}^{N} F_i R_i|V_i|}{\sum_{i=1}^{N} F_i 
R_i\sqrt{V_i^2+\sigma_i^2}},
\end{equation}
where $R_\mathrm{i}$ is the projected radius, $V_\mathrm{i}$ is the 
line-of-sight velocity, $F_\mathrm{i}$ is projected flux (represented here by 
the mass) and $\sigma_\mathrm{i}$ is the line-of-sight velocity dispersion. The
sum is taken over $N$ Voronoi grid cells within the projected radius, so that the radial
$\lambda_{\rm{R}}$-profile is cumulative.

The $\lambda_{\rm{R}}$-parameter can be used to divide elliptical galaxies into fast and slowly 
rotating galaxies with the threshold often set at $\lambda_{\rm{Re}}=0.1$ measured at the effective radius.
A revised threshold has also been
introduced by \citet{2011MNRAS.414..888E}, who suggest in addition to take into account also the shape of 
the galaxies when classifying ETGs by using a criterion $\lambda_{\rm{Re}}=0.31\sqrt{\epsilon_{\rm{e}}}$
where $\epsilon_{\rm{e}}$ is the ellipticity at the effective radius.
An even further refinement is provided by \citet{2016ARA&A..54..597C} who advocate a 
criterion with the threshold set at $\lambda_{\rm{Re}}=0.08+\epsilon_{\rm{e}} /4$
for $\epsilon_{\rm{e}}\le0.4$, which provides a better classification for particularly round ETGs.
Finally, the rotational properties of ETGs can potentially also be used to infer the formation 
mechanism of these galaxies \citep{2009ApJS..182..216K}, and the value of 
$\lambda_{\rm{Re}}$ usually correlates with the kinematic features of the 
galaxies \citep{2007MNRAS.379..418C, 2013MNRAS.433.2812K}.

The radial distribution of stellar $\lambda_{\rm{R}}$ in $0.5$ Gyr intervals during the evolution of the Antennae 
remnant is shown in the left panel of Fig. \ref{fig:lambda_R}.
At the $r$-band effective radius of $r_{\rm{e}}\approx1.6$ kpc we get a value 
$\lambda_{\rm{Re}}\approx0.11$ for the $1$ Gyr old remnant, just above the thresholds for slow rotation. 
Slow rotators can be produced in binary merger simulations when one of the progenitors begins with retrograde
rotation with respect to the orbital motion \citep{2011MNRAS.416.1654B}. Even though the Antennae progenitors are not counter-rotating, 
they are both on quite inclined orbits with $i=60^{\circ}$ and thus result in a mildly rotating merger remnant. 
The radial rotation profiles in Fig. \ref{fig:lambda_R} are
monotonically increasing until they reach plateaus at $r>0.5r_{\rm{e}}$.
The radial angular momentum profiles remain fairly similar with only a slight steepening in 
the inner regions during the evolution of the remnant.
The value of $\lambda_{\rm{R}}$ evolves 
upwards up to a value of $\sim0.15$ as the remnant matures to the age of $3$ Gyr. This
increase of angular momentum is due to the star-forming
disc in the central region, as was also seen with the mean velocity distribution.
If we limit our study to the the evolution of the angular momentum in the stars already present in the $1$ Gyr old remnant,
the $\lambda_{\rm{R}}$-profile does not evolve during the final $2$ Gyr.

The right panel of Fig. \ref{fig:lambda_R} shows the $\lambda_{\rm{Re}}$ values
for the evolving remnant as a function of the respective ellipticity, along with the 
threshold $\lambda_{\rm{Re}}$ values to aide the classification of the remnant. The curves in the right hand panel of
Fig. \ref{fig:lambda_R} show the difference between the traditional threshold $\lambda_{\rm{Re}}=0.1$ and the revised
$\lambda_{\rm{Re}}=0.31\sqrt{\epsilon_{\rm{e}}}$ threshold, and the curves demonstrate how the classification of the
Antennae remnant as a fast or a slow rotator actually depends on the chosen threshold.
Even though the remnant has a fairly low angular momentum for the duration of its evolution, it would still
always be classified as a fast rotator using the traditional old criterion.
However, the low $\lambda_{\rm{Re}}$ values combined with an increasing projected ellipticity drives
the remnant towards and below the revised \citet{2011MNRAS.414..888E} threshold as the remnant ages. Thus in this case 
the remnant would be classified as a fast rotator only before it reaches approximately an age of $1$ Gyr.

In addition, we also show the $\lambda_{\rm{Re}}-\epsilon_{\rm{e}}$ values for 
all the ETGs in the \atlas\ sample in the right hand panel of Fig. \ref{fig:lambda_R}, obtained from \citet{2011MNRAS.414..888E}.
As the scatter in the observed $\lambda_{\rm{Re}}-\epsilon_{\rm{e}}$ values of these ETGs is rather large, 
correlations with additional observables have also been studied to more consistently
classify ETGs. The value of $\lambda_{\rm{Re}}$ 
is commonly compared with the stellar mass. 
Massive ETGs ($M_{\mathrm{dyn}}>10^{11.5}$ M$_{\sun}$) in the \atlas\ sample
mainly populate the low-$\epsilon_{\rm{e}}$, low-$\lambda_{\rm{Re}}$ part of the $\lambda_{\rm{Re}}$-$\epsilon_{\rm{e}}$
diagram which coincides with the slow rotator definition
given in \citet{2016ARA&A..54..597C}. Meanwhile, the lower mass ETGs similar to the Antennae remnant
fill a much larger area in the $\lambda_{\rm{Re}}$-$\epsilon_{\rm{e}}$ diagram \citep{2011MNRAS.414..888E}, including the region 
into which the Antennae remnant evolves.

\section{Finding the local counterparts}\label{section:atlas_counterpart}

\begin{table*}
 \caption{Observed properties of the best look-alikes of the Antennae merger remnant 
 in the \atlas\ sample. Properties derived from the mock-images are provided both with and without the resampling of the young stars.}
 \label{tab:atlas_lookalikes}
 \begin{minipage}{26cm}
 \begin{tabular}{l cccc c}
  \hline
  Property & $1$ Gyr old remnant & NGC 3226 & NGC 3379 & NGC 4494 & Most similar\\ [2pt]
  \hline
  
  $M_*$\footnote{Total stellar mass for the simulated remnant, $M_*\approx L\times (M/L)$ for the \atlas\ galaxies (see text for details) in units of $10^{10}$ M$_{\sun}$} 
  & $10.42$ & $9.84$ & $8.22$ &  $9.84$ & NGC 3226/NGC 4494 \\[2pt]
  
  $\epsilon_{\rm{e}}$\footnote{Ellipticity at $r_{\rm{e}}$}
  & $0.12$ &  $0.168$ & $0.13$ & $0.173$ & NGC 3379 \\[2pt]
  
  $r_{\rm{e}}$\footnote{Single component fit effective radius at $r$-band in kpc, assuming the distance estimates given in \citet{2011MNRAS.413..813C}.}
  & $1.61/1.97$ & $7.26$ & $2.48$ & $3.62$ & NGC 3379 \\ [2pt]
  
  $\mu_{\rm{e}}$\footnote{Single component fit surface brightness at effective radius in mag$/$arcsec$^2$ at $r$-band.}
  & $19.81/20.26$ & $22.95$ & $20.83$ & $21.03$ & NGC 3379 \\ [2pt]
  
  $n$ \footnote{Single component fit S\'{e}rsic index at $r$-band.}
  & $7.07/5.77$ & $4.6$ & $5.3$ & $3.4$ & NGC 3379 \\ [2pt]
  \hline
  
  $\lambda_{\rm{Re}}$ \footnote{Specific angular momentum at $r$-band effective radius}
  & $0.11$ & $0.251$ & $0.157$ & $0.212$ & NGC 3379 \\ [2pt]
  
  $\left\langle V\right\rangle$\footnote{Velocities in km$/$s} 
  & $-53/+61$ & $-81/+90$ & $-90/+93$ &  $-94/+87$ & NGC 3226 \\[2pt]
  
  $\sigma$ & $100/268$ & $97/216$ & $113/395$ &  $96/195$ & NGC 3226 \\[2pt]
  
  Kinematic features 
  & 2 $\left\langle V\right\rangle$ maxima & - & - &  2 $\left\langle V\right\rangle$ maxima & NGC 4494 \\[2pt]
  
  \hline
 \end{tabular}
 \end{minipage}
\end{table*}

\subsection{Local  \atlas\ counterparts}

We have shown in Sections \ref{section:photometric_properties} and 
\ref{section:kinemetry} that the Antennae merger remnant will evolve into an 
early-type galaxy with properties in good agreement with local observed 
ETGs. We now search the \atlas\ sample for the most promising look-alikes
in the local universe, as a comparison for the $1$ Gyr old Antennae merger remnant.

The \atlas\ galaxies are provided with $T$-types \citep{1959HDP....53..275D, 1963ApJS....8...31D} 
and classified in the survey as either S0 if $T>-3.5$
or ellipticals if $T\le -3.5$ \citep{2011MNRAS.413..813C}.
We begin by considering only galaxies which have originally been classified as ellipticals.
To narrow down the search we take the luminosities and mass-to-light ratios ($M/L$)
from \citet{2013MNRAS.432.1709C} obtained with the Jeans Anisotropic Modelling (JAM)\footnote{
JAM expands the widely used Jeans equations to account for anisotropic velocity distributions.
The method uses observed surface brightness photometry and 
integral field observations to extract for example the
inclinations and mass-to-light ratios of galaxies with multiple kinematic components and spatially 
varying anisotropies.} method \citep{2008MNRAS.390...71C}
and identify ETGs with masses similar to the stellar mass of the remnant, $M_*=10.4\times10^{10}$ M$_{\sun}$.
The stellar mass estimates for the \atlas\ galaxies are obtained from \citet{2013MNRAS.432.1709C} as a function of the 
modelled luminosity $L$ in the $r$-band as
$M_*\approx L\times (M/L)$, assuming that the modelled
$M/L$ is representative of the luminous matter within the inner 
galaxy where contribution from dark matter is assumed to be relatively low. 

There are a few elliptical galaxies within a stellar mass range of $(10\pm 2)\times10^{10}$ M$_{\sun}$.
The final selection of candidates is done based on the observed kinematic structure:
many ellipticals with suitable masses and ellipticities have clearly different
rotational or kinematic properties and we omit those galaxies from our sample. 
Finally three ellipticals, NGC 3226, NGC 3379 and NGC 4494 
remain. These galaxies are classified in the \atlas\ sample as having similar
kinematic properties to our remnant with regular rotation, fairly low $\lambda_{\rm{Re}}$-values and 
either no special kinematic features as is the case with NGC 3226 and NGC 3379 or a aligned double 
maxima in $\left\langle V\right\rangle$ as is the case for NGC 4494 \citep{2011MNRAS.414.2923K}. 

The characteristic values describing the kinematic structure and surface brightness profiles 
used in the selection of the best candidates are listed for the three
galaxies in Table \ref{tab:atlas_lookalikes} along with
the respective values for the $1$ Gyr old Antennae remnant. The final column in Table \ref{tab:atlas_lookalikes}
indicates which of the galaxies is identified as the best look-alike in each category.

\begin{figure}
\centering
\includegraphics[width=\columnwidth]{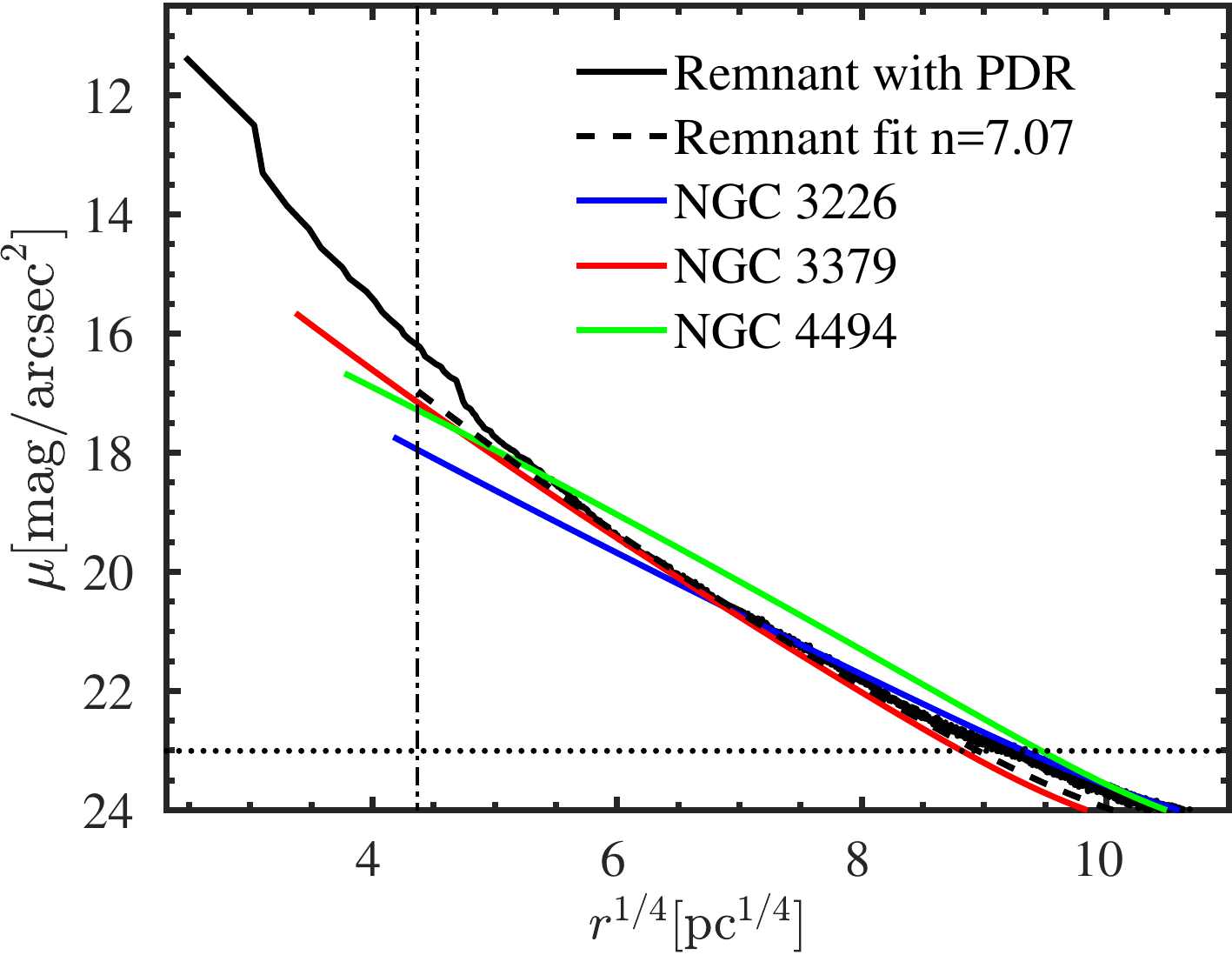}
    \caption{$r$-band surface brightness profile of the $1$ Gyr old 
Antennae remnant (solid black, including the resampling of the young stars) 
    and its best-fit S\'{e}rsic profile (dashed black),
    compared to the best fit profiles of the selected \atlas\ look-alikes (coloured lines). 
    The dotted line shows the mean SDSS sky value indicating a typical outer limit for the 
light profile fitting of \atlas\ galaxies,
    and the dot dashed line shows the inner mask limit of $2.5$ arcsec used
    in the fitting of light profiles of the Antennae remnant. The 
    S\'{e}rsic profiles are only shown from the respective $2.5$ arcsec mask outwards.
    NGC 3379 has a S\'{e}rsic profile very similar to the best-fit profile of the Antennae remnant.
   }
    \label{fig:atlas_lookalike_mu}
\end{figure}

\begin{figure*}
    \includegraphics[width=0.7\columnwidth, angle=90]{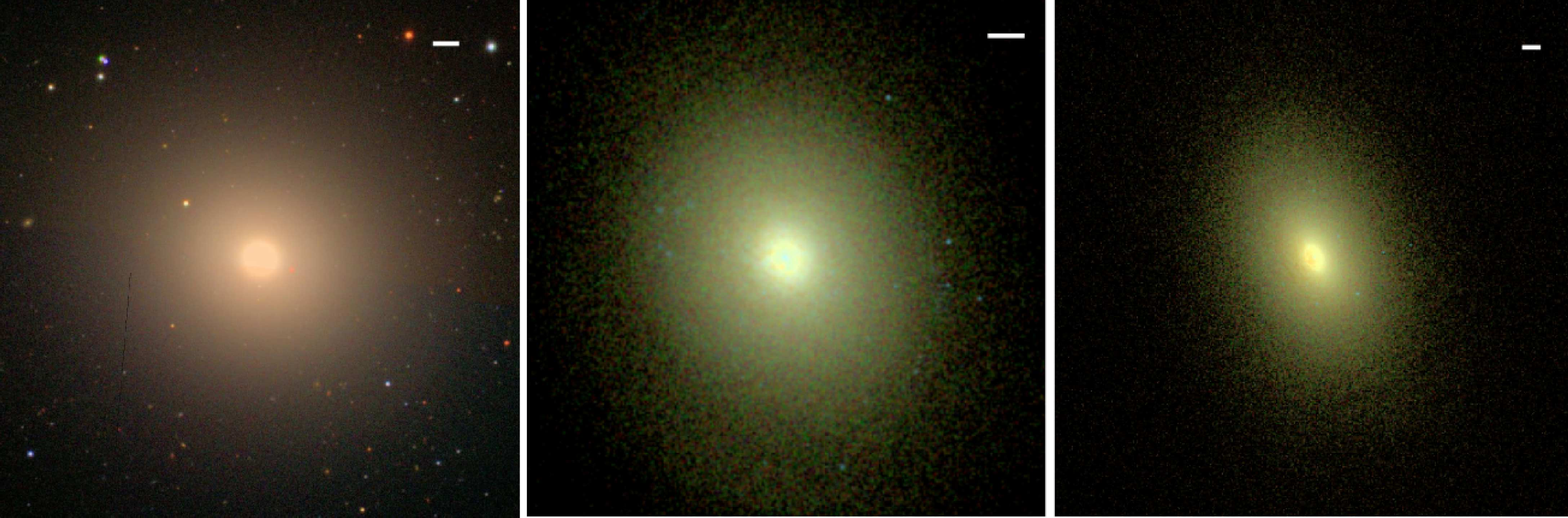}
    \caption{Left: $g$, $r$ and $i$ band colour-composite image of the \atlas\ galaxy NGC 3379 in SDSS DR7 obtained from the SDSS SkyServer. 
    The image spans $10$ times the effective radii provided in \citet{2011MNRAS.413..813C}. 
    Middle and right: $g$-, $r$-, and $i$ colour composite image of the remnant at $1$ Gyr 
    and $3$ Gyr after the final coalescence of the progenitor galaxies, produced with similar photometric
    stretch and asinh-scaling as the \atlas\ image in the left panel.
    The images are produced at the SDSS-resolution of $0.396$ arcsec$/$pix with \skirt\ (see Sec. 
\ref{section:post_processing}) including the resampling of the young stellar particles.
The images span roughly $10$ times the $r$-band effective radius of the respective galaxies and the bars show
    a scale of $1$ kpc.
    }
    \label{fig:remnant_v_atlas}
\end{figure*}

\subsection{Photometric properties}

In Fig. \ref{fig:atlas_lookalike_mu} we show the best fit single component surface brightness profiles for the
three ellipticals obtained from \citet{2013MNRAS.432.1768K}, together with the profile 
of the $1$ Gyr old Antennae remnant (including the resampling of young stars) and its best-fit S\'{e}rsic profile.
The best fit parameters for the single component S\'{e}rsic profiles of the \atlas\ ellipticals\footnote{
The one dimensional best fit to the surface brightness profiles has been obtained by azimuthally averaging the light distribution in the images 
along the best fit ellipses found using the \kinemetry\ code \citep{2006MNRAS.366..787K} and by minimising the
variations of the fitted intensity along the ideal ellipses at each radius.} are summarised in 
Table \ref{tab:atlas_lookalikes}. 
All the chosen candidates span a very similar range of surface brightnesses.
This confirms that the photometric profile of the Antennae remnant is in good agreement with observations 
of the local ETG population with similar masses.  Although the observed galaxies could have extra-light in the centres, the light profiles of the \atlas\ 
galaxies by applying the mask discussed in Sec. \ref{section:photometric_properties} and are
therefore compared to our remnant only in the fitted radial range.

The surface brightness profile of 
NGC 3379 shows a striking similarity with the mock-profile of our remnant in Fig. \ref{fig:atlas_lookalike_mu}, and it
also has the smallest effective radius and largest effective brightness among the ETG candidates.
Based on the profiles in Fig. \ref{fig:atlas_lookalike_mu}, we show in Fig. \ref{fig:remnant_v_atlas} the $g$, $r$ and $i$ 
band composite image of NGC 3379, obtained from the SDSS
SkyServer\footnote{http://skyserver.sdss.org/dr7/en/}, as the visual look-alike of the Antennae remnant.
The observed image can be directly compared to the $g$, $r$ and $i$ composites of the $1$ Gyr and $3$ Gyr old Antennae remnant in Fig. 
\ref{fig:remnant_v_atlas}, constructed with a similar technique as used for the SDSS composite images, showing the same
extent of $10$ effective radii. The effective radius of the Antennae remnant
increases by a factor of $\sim1.5$ as the galaxy evolves, thus the $3$ Gyr image spans a larger region. The image of NGC 3379 has a better spatial resolution
as the distance of NGC 3379 has been approximated to be 
$10.3$ Mpc, $\sim 3$ times closer than our adopted distance for the Antennae.

\begin{figure*}
\centering
\includegraphics[width=0.9\textwidth]{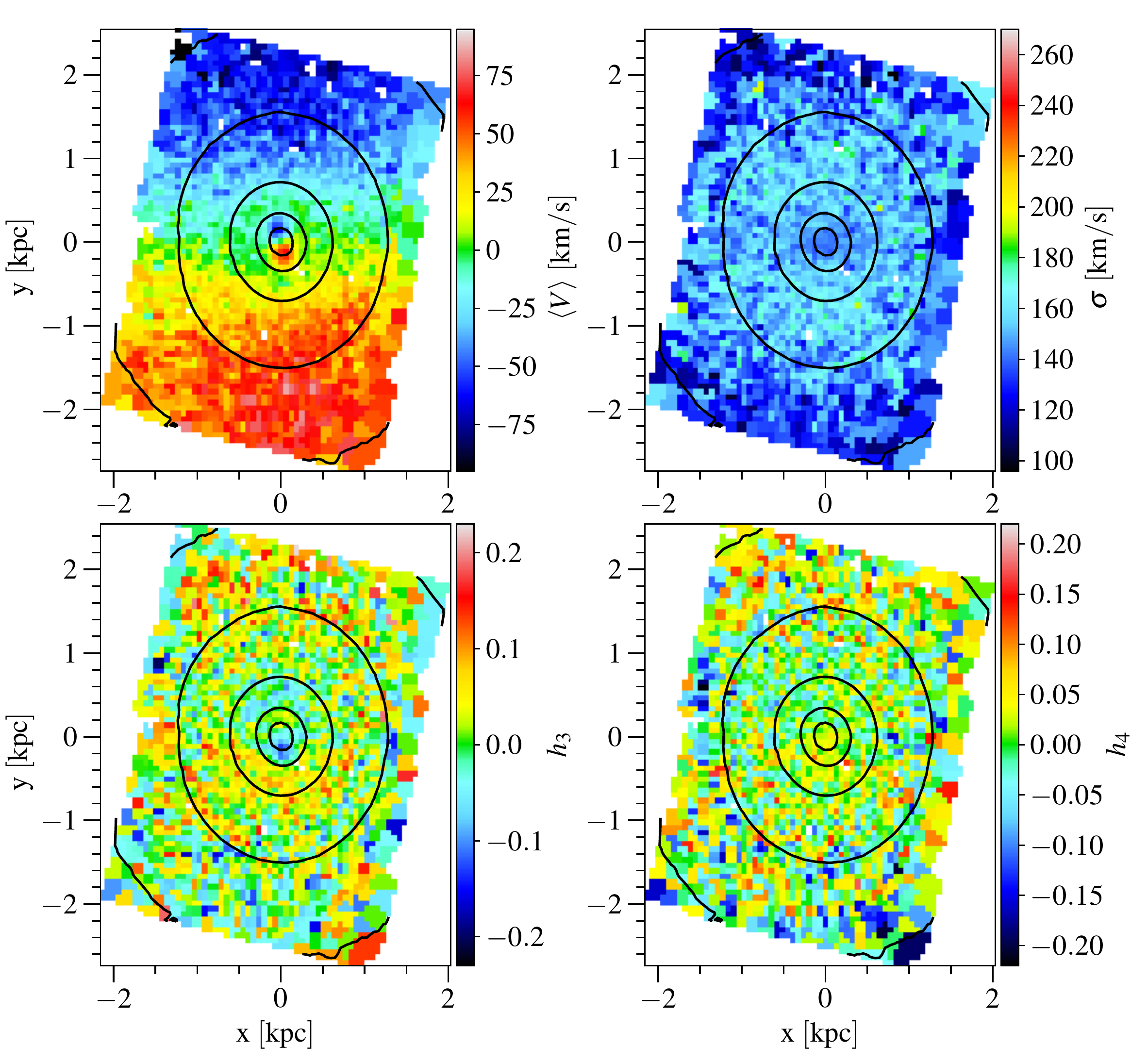}
    \caption{Same as in Fig. \ref{fig:velocity_distribution} for the \atlas\ galaxy NGC 4494 \citep{2011MNRAS.413..813C, 2011MNRAS.414.2923K}
    showing the mean line-of-sight velocity $\langle V \rangle$ (top left), velocity dispersion $\sigma$ (top right), 
    and the higher order Gauss-Hermite coefficients $h_3$ (bottom left) and $h_4$ (bottom right).
    The contours show the distribution of flux in $1$ mag intervals and the LOSVD ranges are the same as in Fig. \ref{fig:velocity_distribution}.}
    \label{fig:atlas_lookalike_velocities}
\end{figure*}

\subsection{Kinematics}

All three candidates have slightly larger $\lambda_{\rm{Re}}$ values compared to the Antennae remnant, which is expected
since for observed galaxies in this mass range the values of $\lambda_{\rm{Re}}$ span the entire range
from $\lambda_{\rm{Re}}=0$ to $\lambda_{\rm{Re}}\rightarrow1$ (see e.g. Fig. 3 in \citealt{2011MNRAS.414..888E}).
Again NGC 3379 provides the most similar value $\lambda_{\rm{Re}}=0.157$ when compared to the $\lambda_{\rm{Re}}=0.12$ value of the Antennae. 
As the ellipticity of NGC 3379 is very similar to our simulated remnant, NGC 3379 is found very
near to our remnant in the right hand panel of Fig. \ref{fig:lambda_R}. NGC 3379 and NGC 4494 
have qualitatively similar radial $\lambda_{\rm{R}}$ profiles as the Antennae remnant, based on measurements of 
$\lambda_{\rm{R}}$ at $r_{\rm{e}}$ and $r_{\rm{e}} /2$ \citep{2011MNRAS.414..888E}. The $\lambda_{\rm{R}}$ values are either smaller or equal at 
the half effective radii compared to the $\lambda_{\rm{R}}$ at the effective radius. NGC 3226, on the
other hand, may have a centrally peaked, radially decreasing $\lambda_{\rm{R}}$ profile
as the $\lambda_{\rm{R}}$ at $r_{e/2}$ is slightly larger than at $r_{e}$.

The range of $\langle V\rangle$ for all the three observed galaxies is slightly larger than for the Antennae while only for 
NGC 3379 the velocity dispersion of the central region exceeds that of the Antennae remnant.
Fig. \ref{fig:atlas_lookalike_velocities}, which can be directly compared to the corresponding Fig. 
\ref{fig:velocity_distribution}, shows the velocity distribution of NGC 4494
where a double $\left\langle V\right\rangle$ maximum feature can be seen within the innermost
$500$ pc. In addition a slight $V/\sigma-h_3$ anticorrelation in the velocity distribution can be observed, 
similar to what is seen in the Antennae remnant. 
The velocity dispersion of NGC 4494, on the other hand, does not show a strong central peak, and when studying the range 
of LOSVD values of the Antennae remnant, NGC 3226 would be a better look-alike (see Table \ref{tab:atlas_lookalikes}).

In conclusion, all of these observed candidates show some indisputable similarities with 
the simulated merger remnant of the Antennae galaxies.
The surface brightness profile of NGC 3379, shown in Fig. \ref{fig:remnant_v_atlas}, 
makes it a good example for the potential future visual appearance of the 
Antennae remnant, while NGC 3226 represents an example of what kinematic observations
of the future remnant might look like. On the other hand NGC 4494 provides an example of the commonly observed double maximum 
feature in the mean velocity distribution which works as a indicator for a central disc structure.

\section{Conclusions}

We have run and analysed a high-resolution hydrodynamical simulation representing the evolution of the 
Antennae galaxy merger (NGC 4038/4039) up to $3.2$ Gyr into the future. 
The use of post-processing methods enabled a direct comparison to observations of both the Antennae system and
local early-type galaxies. Full radiative transfer modelling from ultra-violet to far-infrared wavelengths has been
performed for the first time on the simulated particle data of an Antennae look-alike.
Based on the spatial infrared SED, we have derived a ratio of $4.5$ for the star formation rate (SFR) in the overlap region
versus the SFR in the nuclei for the simulated best-match Antennae, a value close to the observed ratio of $\sim 3$--$6$.
In total we obtain an SFR of $23.6$ M$_{\sun}/$yr for the simulated Antennae, which
slightly exceeds the observed value of $22.2$ M$_{\sun}/$yr.
The match between the simulations and the observations could primarily be improved by tuning the initial conditions of the
NGC 4039 progenitor in order to produce a lower star formation level at the time of the best match.

The refined disc galaxy initial conditions with initial metallicity and age distributions motivated by observations of
the Milky Way have enabled the first simulation of the spatial stellar metallicity distribution in the Antennae system.
Our metallicity values, which mostly range from solar to slightly supersolar, are in good agreement with the metallicities
of observed young stellar clusters located in off-nuclear sites \citet{2009ApJ...701..607B,2015ApJ...812..160L}. However, the simulated
metallicity in the nuclei of the still distinct galactic discs were at most $0.7$ Z$_{\sun}$ higher than observed. 
A more observationally motivated analysis would therefore be warranted, which would use metallicity estimates based on spectral line indices 
and an observationally equivalent selection of stellar clusters based on cluster ages.

The photometric and kinematic properties of the  $1$ Gyr old Antennae merger remnant has been scrutinised in detail, with the results 
compared to observational surveys of ETGs, in particular the \atlas\ survey. The analysis of the shape of the remnant reveals a fairly spherical, isotropic
galaxy with intrinsic axis ratios predominantly in the range $0.8$--$0.9$ for the entire radial range. 
The observed shape, including dust obscuration and projection effects, calculated by fitting a S\'{e}rsic profile to the
surface brightness profile of the post-processed SDSS equivalent $r$-band image, results in an ellipticity at $r_{\rm{e}}$
of $\epsilon_{\rm{e}}=0.122$. This is only slightly larger than the respective intrinsic value of $\epsilon_{\rm{e}}\approx 0.09$.  
The best fit light profile has a S\'{e}rsic index of $n=7.07$, which is toward the upper end of typical values for local ETGs.
The $r$-band effective radius $r_{\rm{e}}=1.6$ kpc of the $1$ Gyr old remnant is relatively small and this is primarily caused by
the nuclear extra-light in the central disc. As a consequence, the central properties within $r_{\rm{e}}$ place the $1$ Gyr old remnant on the observed fundamental plane of ETGs,
but towards the bright and compact end of the observed range. During the evolution of the merger remnant the effective radius increases
as the central light fades after most of the star formation in the central gaseous disc is extinguished. 

The negative intrinsic radial metallicity abundance gradient of the stellar particles
flattens from the initial $0.06$ dex$/$kpc value to $0.028$ dex$/$kpc in the $1$ Gyr old remnant, while the gaseous oxygen 
abundance shows a more centrally peaked 
distribution with a flatter gradient in the outer regions. The resulting radially decreasing abundances agree with the overall
 trends in metallicities of observed ETGs.
The central oxygen abundances of both stars and gas are enhanced with respect to the initial conditions due to stellar enrichment and star formation
from enriched gas.

After coalescence, the young remnant shows observational properties reminiscent of later-type
galaxies, as the apparent $u-r$ colour is initially quite blue. The remnant shows an asymptotically decreasing 
amount of star formation in its central region 
after the merger induced starburst has faded. As a result the young stars initially prevent the remnant
from crossing in the $u-r$ vs. $M_r$ colour-magnitude diagram from the blue cloud 
to the ETG dominated red sequence. Only after $2.5$ Gyr of 
isolated evolution can the remnant be identified as a red ETG.
In a similar manner, some local ETGs with colours bluer than expected in the \atlas\ sample have been
identified as possibly transitioning objects between the blue cloud and the red sequence.

The spatial kinematic analysis of the stellar particles of the Antennae remnant shows a LOSVD with clearly ordered rotation. 
The double maximum feature in the $\langle V \rangle$ distribution
of the remnant is a common feature, also seen in local \atlas\ ETGs. 
The central star-forming disc is manifested in the observational 
features as the double maxima in $\langle V \rangle$ and a clearly anticorrelated $\langle V \rangle/\sigma$--$h_3$ distribution, which are a
natural consequence of a gas-rich major merger.

The fairly slow rotational velocity in the evolving remnant is a consequence of the orbital configuration of the Antennae galaxies: the
progenitors merge with highly inclined disc orientations, with $i=60^{\circ}$ with respect to the direction of the orbital angular momentum. 
This results in a fairly low angular momentum in the final merger remnant.
The specific angular momentum $\lambda_{\rm{Re}}$ at the effective radius is used to quantify the rotation, and we obtain values
increasing from $\lambda_{\rm{Re}}=0.11$ to $\lambda_{\rm{Re}}=0.14$ during the final $2$ Gyr of evolution. 
The revised threshold of $\lambda_{\rm{Re}}=0.31\sqrt{\epsilon_{\rm{e}}}$ for slow versus fast rotation
would classify the remnant as a slow rotator for remnant ages $> 1$ Gyr, whereas
the traditional threshold $\lambda_{\rm{Re}}=0.1$ would classify the remnant as a fast rotator throughout its evolution.

Finally, we identified three local ETGs in the \atlas\ sample with similar masses and photometric and kinematic properties reminiscent of the values
derived for the $1$ Gyr old Antennae merger remnant. A visual counterpart with a similar surface brightness profile is provided by NGC 3379, whereas
NGC 3226 has LOSVD properties in a similar range, and NGC 4494 shows the same kind of double maximum feature
as was found for the Antennae remnant. 
Based on the kinematic features mainly dominated by the presence of the nuclear disc the Antennae remnant would be classified 
in the second most common class of local  \atlas\ ETGs. This class includes 
$14\%$ of the \atlas\ ETG sample, exhibiting ordered rotation and a special, yet not entirely decoupled, kinematic feature. 
Overall, we conclude that the famous Antennae merger will evolve into an early-type galaxy with properties very similar to the observed 
properties of the general local early-type galaxy population.

\section*{Acknowledgements}

The authors would like to thank  Pauli Pihajoki and Till Sawala for helpful discussions and comments on the manuscript, and Milena Pawlik
for help in producing the SDSS equivalent images of the simulated outputs. The numerical simulations were
performed on facilities hosted by the CSC-IT Center for Science in Espoo, Finland.
N.L.,  P.H.J and A.R. acknowledge support from the MPA Garching Visitor Programme.
N.L. is supported by the Jenny \& Antti Wihuri Foundation and the Doctoral Programme
in Particle Physics and Universe Sciences at the University of Helsinki.
N.L., P.H.J. and A.R. acknowledge the support of the Academy of Finland grant 274931.
A.R. is funded by the doctoral program of Particle Physics and Universe Sciences at the University of Helsinki




\bibliographystyle{mnras}
\bibliography{references} 






\bsp	
\label{lastpage}
\end{document}